\documentclass[12pt,dvips]{article}
\usepackage{rotating}
\usepackage{axodraw}
\usepackage{epsfig}
\usepackage{color}
\usepackage{cite}
\definecolor{GreenYellow}   {cmyk}{0.15,0,0.69,0}
\definecolor{Yellow}        {cmyk}{0,0,1,0}
\definecolor{Goldenrod}     {cmyk}{0,0.10,0.84,0}
\definecolor{Dandelion}     {cmyk}{0,0.29,0.84,0}
\definecolor{Apricot}       {cmyk}{0,0.32,0.52,0}
\definecolor{Peach}         {cmyk}{0,0.50,0.70,0}
\definecolor{Melon}         {cmyk}{0,0.46,0.50,0}
\definecolor{YellowOrange}  {cmyk}{0,0.42,1,0}
\definecolor{Orange}        {cmyk}{0,0.61,0.87,0}
\definecolor{BurntOrange}   {cmyk}{0,0.51,1,0}
\definecolor{Bittersweet}   {cmyk}{0,0.75,1,0.24}
\definecolor{RedOrange}     {cmyk}{0,0.77,0.87,0}
\definecolor{Mahogany}      {cmyk}{0,0.85,0.87,0.35}
\definecolor{Maroon}        {cmyk}{0,0.87,0.68,0.32}
\definecolor{BrickRed}      {cmyk}{0,0.89,0.94,0.28}
\definecolor{Red}           {cmyk}{0,1,1,0}
\definecolor{OrangeRed}     {cmyk}{0,1,0.50,0}
\definecolor{RubineRed}     {cmyk}{0,1,0.13,0}
\definecolor{WildStrawberry}{cmyk}{0,0.96,0.39,0}
\definecolor{Salmon}        {cmyk}{0,0.53,0.38,0}
\definecolor{CarnationPink} {cmyk}{0,0.63,0,0}
\definecolor{Magenta}       {cmyk}{0,1,0,0}
\definecolor{VioletRed}     {cmyk}{0,0.81,0,0}
\definecolor{Rhodamine}     {cmyk}{0,0.82,0,0}
\definecolor{Mulberry}      {cmyk}{0.34,0.90,0,0.02}
\definecolor{RedViolet}     {cmyk}{0.07,0.90,0,0.34}
\definecolor{Fuchsia}       {cmyk}{0.47,0.91,0,0.08}
\definecolor{Lavender}      {cmyk}{0,0.48,0,0}
\definecolor{Thistle}       {cmyk}{0.12,0.59,0,0}
\definecolor{Orchid}        {cmyk}{0.32,0.64,0,0}
\definecolor{DarkOrchid}    {cmyk}{0.40,0.80,0.20,0}
\definecolor{Purple}        {cmyk}{0.45,0.86,0,0}
\definecolor{Plum}          {cmyk}{0.50,1,0,0}
\definecolor{Violet}        {cmyk}{0.79,0.88,0,0}
\definecolor{RoyalPurple}   {cmyk}{0.75,0.90,0,0}
\definecolor{BlueViolet}    {cmyk}{0.86,0.91,0,0.04}
\definecolor{Periwinkle}    {cmyk}{0.57,0.55,0,0}
\definecolor{CadetBlue}     {cmyk}{0.62,0.57,0.23,0}
\definecolor{CornflowerBlue}{cmyk}{0.65,0.13,0,0}
\definecolor{MidnightBlue}  {cmyk}{0.98,0.13,0,0.43}
\definecolor{NavyBlue}      {cmyk}{0.94,0.54,0,0}
\definecolor{RoyalBlue}     {cmyk}{1,0.50,0,0}
\definecolor{Blue}          {cmyk}{1,1,0,0}
\definecolor{Cerulean}      {cmyk}{0.94,0.11,0,0}
\definecolor{Cyan}          {cmyk}{1,0,0,0}
\definecolor{ProcessBlue}   {cmyk}{0.96,0,0,0}
\definecolor{SkyBlue}       {cmyk}{0.62,0,0.12,0}
\definecolor{Turquoise}     {cmyk}{0.85,0,0.20,0}
\definecolor{TealBlue}      {cmyk}{0.86,0,0.34,0.02}
\definecolor{Aquamarine}    {cmyk}{0.82,0,0.30,0}
\definecolor{BlueGreen}     {cmyk}{0.85,0,0.33,0}
\definecolor{Emerald}       {cmyk}{1,0,0.50,0}
\definecolor{JungleGreen}   {cmyk}{0.99,0,0.52,0}
\definecolor{SeaGreen}      {cmyk}{0.69,0,0.50,0}
\definecolor{Green}         {cmyk}{1,0,1,0}
\definecolor{ForestGreen}   {cmyk}{0.91,0,0.88,0.12}
\definecolor{PineGreen}     {cmyk}{0.92,0,0.59,0.25}
\definecolor{LimeGreen}     {cmyk}{0.50,0,1,0}
\definecolor{YellowGreen}   {cmyk}{0.44,0,0.74,0}
\definecolor{SpringGreen}   {cmyk}{0.26,0,0.76,0}
\definecolor{OliveGreen}    {cmyk}{0.64,0,0.95,0.40}
\definecolor{RawSienna}     {cmyk}{0,0.72,1,0.45}
\definecolor{Sepia}         {cmyk}{0,0.83,1,0.70}
\definecolor{Brown}         {cmyk}{0,0.81,1,0.60}
\definecolor{Tan}           {cmyk}{0.14,0.42,0.56,0}
\definecolor{Gray}          {cmyk}{0,0,0,0.50}
\definecolor{Black}         {cmyk}{0,0,0,1}
\definecolor{White}         {cmyk}{0,0,0,0}

\newcommand{\imag}{\Im {\rm m}}
\newcommand{\real}{\Re {\rm e}}

\newcommand{\lsim}{\raisebox{-0.13cm}{~\shortstack{$<$ \\[-0.07cm] $\sim$}}~}
\newcommand{\gsim}{\raisebox{-0.13cm}{~\shortstack{$>$ \\[-0.07cm] $\sim$}}~}
\newcommand{\bra}[1]{\langle #1|}
\newcommand{\ket}[1]{|#1\rangle}

\begin{document}

\def\thefootnote{\fnsymbol{footnote}}

\begin{flushright}
{\tt KEK-TH-1203}, {\tt FERMILAB-PUB-07-651-T}\\
{\tt CERN-PH-TH/2007-258}, {\tt MAN/HEP/2007/43 }\\
{\tt ANL-HEP-PR-07-105}, {\tt EFI-07-38}\\
{\tt arXiv:0712.2360} \\
December 2007
\end{flushright}

\begin{center}
{\bf {\LARGE
{\color{Red}CP}{\color{Blue}super}{\color{OliveGreen}H}{\color{Gray}2.0}:}\\
{\large an Improved Computational Tool for Higgs Phenomenology\\
in the MSSM with Explicit CP Violation}
}
\end{center}

\medskip

\begin{center}{\large
J.~S.~Lee$^{a,b}$,
M.~Carena$^c$,
J.~Ellis$^d$,
A.~Pilaftsis$^e$
and C.~E.~M.~Wagner$^{f,g}$}
\end{center}

\begin{center}
{\em $^a$Theory Group, KEK, Oho 1-1 Tsukuba, 305-0801, Japan}\\[0.2cm]
{\em $^b$Department of Physics, National Central University, Chung-Li, Taiwan
32054}\\[0.2cm]
{\em $^c$Fermilab, P.O. Box 500, Batavia IL 60510, U.S.A.}\\[0.2cm]
{\em $^d$Theory Division, CERN, CH-1211 Geneva 23, Switzerland}\\[0.2cm]
{\em $^e$School of Physics and Astronomy, University of Manchester,}\\
{\em Manchester M13 9PL, United Kingdom}\\[0.2cm]
{\em $^f$HEP Division, Argonne National Laboratory,
9700 Cass Ave., Argonne, IL 60439, USA}\\[0.2cm]
{\em $^g$Enrico Fermi Institute, Univ. of Chicago, 5640
Ellis Ave., Chicago, IL 60637, USA}
\end{center}

\bigskip

\centerline{\bf ABSTRACT}
\medskip\noindent  We describe  the Fortran
code  {\tt CPsuperH2.0}, which contains several improvements and extensions of
its predecessor {\tt CPsuperH}. It implements improved calculations of the
Higgs-boson pole masses, notably a full treatment of the $4 \times 4$
neutral Higgs propagator matrix including the Goldstone boson and a more complete
treatment of threshold effects in self-energies and Yukawa couplings,
improved treatments of two-body Higgs decays, some important three-body
decays, and two-loop Higgs-mediated contributions to electric dipole moments.
{\tt CPsuperH2.0} also implements an integrated treatment of several $B$-meson
observables, including the branching ratios of $B_s \to \mu^+ \mu^-$,
$B_d \to \tau^+ \tau^-$, $B_u \to \tau \nu$, $B \to X_s \gamma$ and the latter's
CP-violating asymmetry ${\cal A}_{\rm CP}$, and the supersymmetric
contributions to the $B^0_{s,d} - {\bar B^0_{s,d}}$ mass differences.
These additions make {\tt CPsuperH2.0} an attractive integrated tool for
analyzing supersymmetric CP and flavour physics as well as searches for new
physics at high-energy colliders such as the Tevatron, LHC and linear colliders.
\footnote{The     program    may    be     obtained    from
{\tt http://www.hep.man.ac.uk/u/jslee/CPsuperH.html}.}

\newpage

\section{Introduction}

With the imminent advent of  the LHC, particle physics experiments are
poised to  explore the TeV energy  range directly for  the first time.
There are several reasons to  expect new physics in this energy range,
such  as  the  origin  of  particle masses  and  electroweak  symmetry
breaking, the  hierarchy problem  and the nature  of dark  matter.  In
parallel  with the  direct  exploration of  the  TeV scale,  precision
experiments at low energies continue to place important constraints on
the  possible  flavour and  CP-violating  structure  of any  TeV-scale
physics. Prominent examples include experiments on $B$ and $K$ mesons,
and  probes of  electric  dipole moments~\cite{Nath}.   It is  clearly
desirable to develop computational tools that can be used to calculate
consistently observables for both  low- and high-energy experiments in
a coherent numerical framework. This is particularly desirable in view
of the possibility that the dominance of matter over antimatter in the
Universe may be due to CP-violating interactions at the TeV scale~\cite{BARYO}.

Supersymmetry is one of the most prominent possibilities for new
TeV-scale physics, and the minimal supersymmetric extension of
the Standard Model (MSSM) provides a natural cold dark matter
candidate as well as stabilizing the electroweak scale and
facilitating the unification of the fundamental interactions. There
are many computational tools available for calculations within
the MSSM. The first to include CP-violating phases was {\tt CPsuperH}~\cite{cpsuperh}
based on the renormalization-group-(RG-)improved effective potential approach.
The Higgs-boson pole-mass shifts are calculated by employing the 
RG-improved diagrammatic approach.
The recent versions of {\tt FeynHiggs}~\cite{feynhiggs} are based the Feynman
diagrammatic approach. There are merits in both approaches and the difference between
two programs may be
attributed to some unknown higher-order corrections.

Some of us have recently published an analysis of several $B$-physics
observables taking into account the most
general set of CP-violating parameters allowed under the assumption
of minimal flavour violation in the supersymmetric
sector~\cite{MCPMFV}. For this purpose we used an
updated and extended computational tool, {\tt CPsuperH2.0},
which we introduce and describe in this paper.

The main new features of {\tt CPsuperH2.0} are its inclusion of a
number of $B$ observables, including the branching ratios of 
$B_s \to \mu^+ \mu^-$, 
$B_d \to \tau^+ \tau^-$, $B_u \to \tau \nu$, $B \to X_s \gamma$ and the latter's
CP-violating asymmetry ${\cal A}_{\rm CP}$, and the supersymmetric
contributions to the $B^0_{s,d} - {\bar B^0_{s,d}}$ mass differences.
In addition, {\tt CPsuperH2.0} includes a more complete treatment
of Higgs-boson pole masses, based on a full treatment of the $4 \times 4$
neutral Higgs propagator matrix including the Goldstone boson and
a more complete treatment of threshold effects in self-energies and Yukawa couplings.
It also includes improved treatments of two-body Higgs decays, some important three-body
decays, and two-loop Higgs-mediated contributions to electric dipole moments.
Therefore, {\tt CPsuperH2.0} provides an essentially complete,
self-contained and consistent computational tool for evaluating flavour and
CP-violating physics at energies up to the TeV scale.

The structure of this paper is as follows. Several updated features of
{\tt CPsuperH2.0} are described in Section~2. In particular, in
Subsection~2.1 we introduce the
improved treatment of Higgs-boson pole masses, and Section~2.2
contains a description of the improvements in the treatment of Higgs decay modes. 
Then, in Section~3 we describe the {\tt CPsuperH2.0} treatment 
of two-loop Higgs effects on electric dipole moments.
The most important new features are described in Section~4, where we
discuss its treatment of $B$ observables. In each Section, we illustrate
in figures some typical results obtained using {\tt CPsuperH2.0}.

\section{Updated Features of {\tt CPsuperH2.0}}
\label{sec:cpsuperh2.0}

It is to be understood that, throughout this paper, we follow the 
notations and conventions defined and adopted in {\tt CPsuperH} 
for the mixing matrices of neutral Higgs bosons, charginos, neutralinos 
and third--generation sfermions, as well as their masses and couplings, etc.
The updates to the original version of {\tt CPsuperH}~\cite{cpsuperh} that are
presented here reflect, in part, feedback from users, as well as extending it
to $B$ observables. 

New common blocks {\tt /HC\_RAUX/} and {\tt /HC\_CAUX/} have been introduced for the general
purpose of storing new numerical outputs which are available in {\tt CPsuperH2.0}: 
\begin{itemize}
\item {\tt COMMON /HC\_RAUX/ RAUX\_H }
\item {\tt COMMON /HC\_CAUX/ CAUX\_H }
\end{itemize}
The two arrays {\tt RAUX\_H} and {\tt CAUX\_H} are {\tt NAUX}$=$999 dimensional and
only parts of them are being used presently as shown in Tables~\ref{tab:raux} and
\ref{tab:caux}. The contents of these two new arrays are explained 
in the corresponding following subsections. These common blocks can also be used by 
users for their specific purposes.

\subsection{Improved Treatment of Higgs-Boson Masses and Propagators}

\begin{figure}[ht]
\hspace{ 0.0cm}
\vspace{-0.5cm}
\centerline{\epsfig{figure=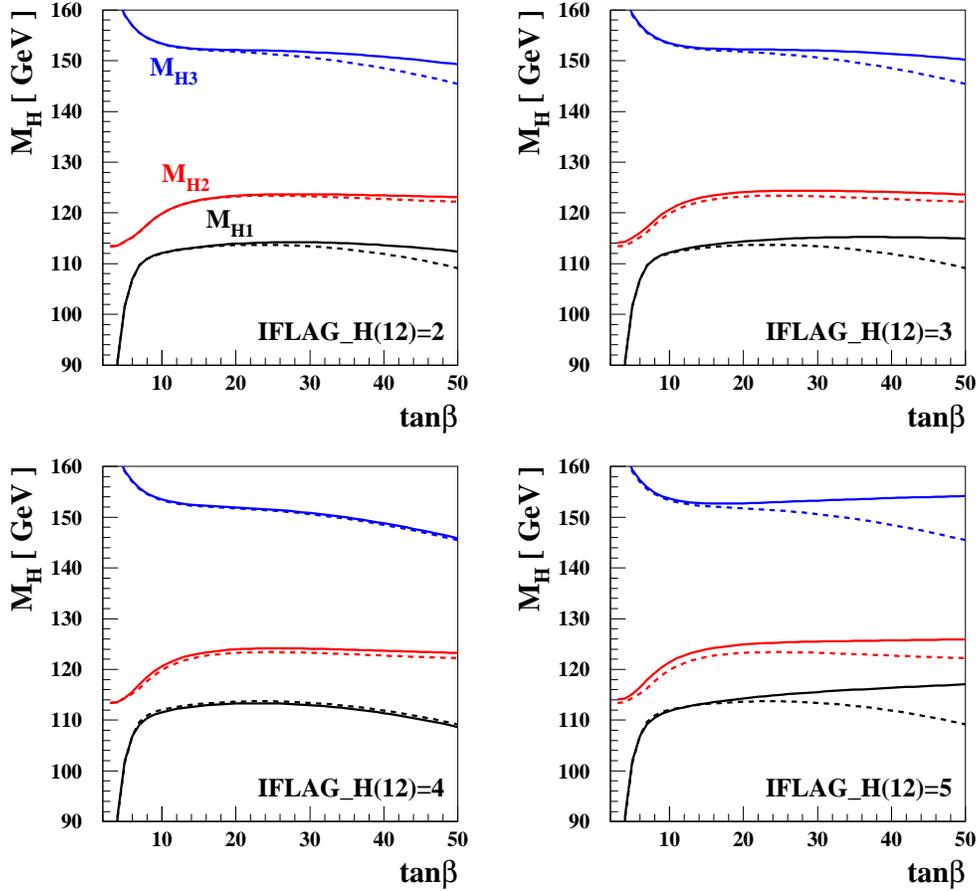,height=13.5cm,width=13.5cm}}
\vspace{-0.5cm}
\caption{\it
The masses of the neutral Higgs bosons as functions of $\tan\beta$ for the CPX scenario
\cite{Carena:2000ks}
taking $\Phi_3=\Phi_{A_{t,b,\tau}}=90^\circ$ in the convention $\Phi_\mu=0$,
$M_{\rm SUSY}=0.5$ TeV, and
the charged Higgs-boson pole mass $M_{H^\pm}=160$ GeV.
In each frame, the dashed line is for the case
{\tt IFLAG\_H(12)}$=1$ and the solid line for other case indicated.
}
\label{fig:mhtb}
\end{figure}
\begin{figure}[htb]
\hspace{ 0.0cm}
\vspace{-0.5cm}
\centerline{\epsfig{figure=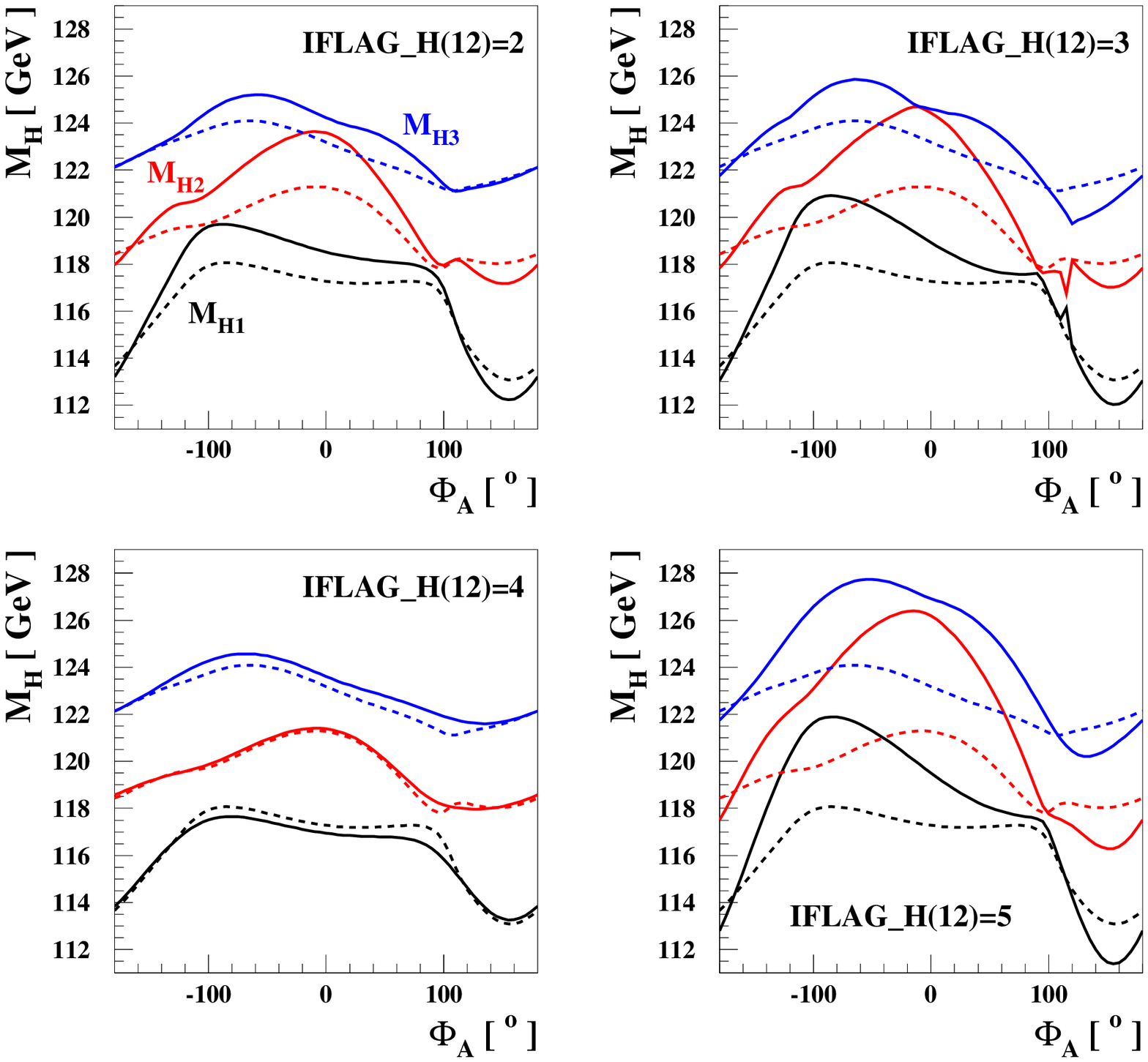,height=14.0cm,width=14.0cm}}
\vspace{-0.5cm}
\caption{\it
The masses of the neutral Higgs bosons as functions of the common phase $\Phi_A$
for the trimixing scenario \cite{Ellis:2004fs}
taking $\Phi_3=-90^\circ$. Specifically, in this scenario,
$\tan\beta=50$ and $M_{H^\pm}=155$ GeV.
The lines are the same as in Fig.~\ref{fig:mhtb}.
}
\label{fig:mhpa}
\end{figure}

In {\tt CPsuperH2.0} we make three main
improvements in the calculation of the Higgs-boson pole masses.

\begin{itemize}
\item  
The finite threshold corrections induced by the exchanges of gluinos
and charginos  have been included
in the top- and bottom-quark self-energies of the
neutral and charged Higgs bosons.
For the explicit expressions
of the self-energies, we refer to Eqs.~(B.14), (B.15), and (B.16) of 
Ref.~\cite{Carena:2001fw}~\footnote{We find that
overall minus signs are missing in the expressions of $\Pi_{11,22}^{P,(c)}(s)$.}.
\item Also included are the threshold corrections to the Yukawa couplings $|h_{t,b}|$ in the
one-loop running quartic couplings, $\lambda_i^{(1)}(Q=m_t^{\rm pole})$ 
with $i=1-4$. For the explicit expressions of $\lambda_i^{(1)}$, we
refer to Eqs.~(3.3)-(3.6) of Ref.~\cite{Carena:2000yi}.
\item An improved 
iterative method has been employed for the calculation of the pole masses.
\end{itemize}
 
As a help in assessing
the improvements in the calculation of Higgs sector,  new flags {\tt IFLAG\_H(12)} and
{\tt IFLAG\_H(60)} have been introduced as follows:

\begin{itemize}
\item {\tt IFLAG\_H(12)}: 
\begin{itemize}
\item {\tt IFLAG\_H(12)}$=1$: Gives the same result as that obtained by the
older version of {\tt CPsuperH}.
\item {\tt IFLAG\_H(12)}$=\!\!2$: Includes only the 
threshold corrections to the neutral and charged Higgs-boson quark self-energies. 
\item {\tt IFLAG\_H(12)}$=3$: Includes only the 
threshold corrections to $\lambda_i^{(1)}$. 
\item {\tt IFLAG\_H(12)}$=4$: Includes only the 
iterative method for the pole masses.
\item {\tt IFLAG\_H(12)}$=5$ or $0$: All the improvements are fully included.
\end{itemize}
\item {\tt IFLAG\_H(60)}$=1$: This is an error message that appears 
when the iterative method for the pole masses fails.
\end{itemize}

\noindent
The improvement in the threshold corrections to the top- and 
bottom-quark Yukawa couplings is important when $\tan\beta$ is large and
the charged Higgs boson is light.
In Figs.~\ref{fig:mhtb} and \ref{fig:mhpa}, we show the pole masses of the neutral Higgs bosons
for the CPX \cite{Carena:2000ks} and trimixing \cite{Ellis:2004fs}
scenarios, respectively, when ${\tt IFLAG\_H(12)}=2$-$5$ as indicated. In each frame, the
old calculation with {\tt IFLAG\_H(12)}$=1$ (dashed line) is also shown for comparison.
 
Finally, {\tt RAUX\_H(1-6)}, {\tt RAUX\_H(10-36)}, and {\tt CAUX\_H(1-2)} are allocated for
numerical information on the Higgs-sector calculation based on a renormalization-group-improved 
diagrammatic approach including dominant higher-order logarithmic and threshold 
corrections~\cite{Carena:2000yi,Carena:2001fw}, see Tables~\ref{tab:raux} and \ref{tab:caux}.

\begin{figure}[htb]
\hspace{ 0.0cm}
\vspace{-0.5cm}
\centerline{\epsfig{figure=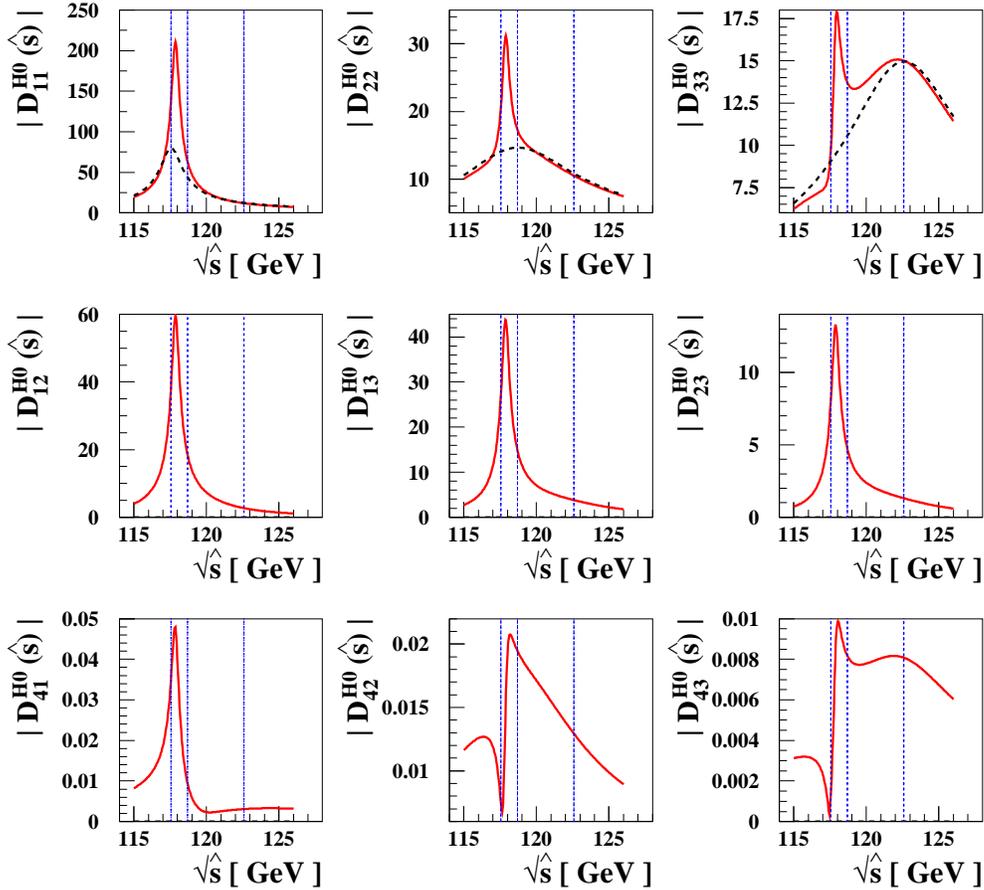,height=13.5cm,width=13.5cm}}
\vspace{-0.5cm}
\caption{\it
The absolute value of each component of the neutral Higgs-boson propagator matrix
$D^{H^0}({\hat{s}})$ with (red solid lines) and without (black dashed lines)
including off-diagonal absorptive parts in the trimixing scenario with
$\Phi_A=-\Phi_3=90^\circ$ and {\tt IFLAG\_H(12)}$=5$. We note that
$|D^{H^0}_{4\,4}({\hat{s}})|=1$.
The three Higgs-boson pole masses are indicated by thin vertical
lines.
}
\vspace{-0.2cm}
\label{fig:dh3}
\end{figure}

In  situations  where  two   or  more  MSSM  Higgs  bosons  contribute
simultaneously to  a process, the transitions  between the Higgs-boson
mass eigenstates need to be  considered before their decays.  For this
reason,  we include  the  {\em   complete}  $4\times  4$-dimensional
propagator     matrix     $D^{H^0}(\hat{s})$     spanned    by     the
basis~$(H_1,H_2,H_3,G^0)$~\cite{APNPB},     including     off-diagonal
absorptive   parts~\cite{Ellis:2004fs}.   The   dimensionless  neutral
Higgs-boson propagator matrix is given by
\begin{eqnarray}
  \label{eq:hprop}
D^{H^0} (\hat{s}) &=&  \nonumber \\
&&\hspace{-2.5cm}
\hat{s}\,
\left(\begin{array}{cccc}
       \hat{s}-M_{H_1}^2+i\imag\widehat{\Pi}_{11}(\hat{s}) &
i\imag\widehat{\Pi}_{12}(\hat{s})&
       i\imag\widehat{\Pi}_{13}(\hat{s}) &
       i\imag\widehat{\Pi}_{14}(\hat{s}) \\
       i\imag\widehat{\Pi}_{21}(\hat{s}) &
\hat{s}-M_{H_2}^2+i\imag\widehat{\Pi}_{22}(\hat{s})&
       i\imag\widehat{\Pi}_{23}(\hat{s}) &
       i\imag\widehat{\Pi}_{24}(\hat{s}) \\
       i\imag\widehat{\Pi}_{31}(\hat{s}) &
i\imag\widehat{\Pi}_{32}(\hat{s}) &
       \hat{s}-M_{H_3}^2+
       i\imag\widehat{\Pi}_{33}(\hat{s}) &
       i\imag\widehat{\Pi}_{34}(\hat{s}) \\
i\imag\widehat{\Pi}_{41}(\hat{s})&
       i\imag\widehat{\Pi}_{42}(\hat{s}) &
       i\imag\widehat{\Pi}_{43}(\hat{s}) &
       \hat{s} +i\imag\widehat{\Pi}_{44}(\hat{s}) 
      \end{array}\right)^{-1} , \nonumber \\
\label{eq:Hprop}
\end{eqnarray}
where $M_{H_{1,2,3}}$  are the  one-loop Higgs-boson pole  masses, and
higher-order   absorptive  effects   on   $M_{H_{1,2,3}}$  have   been
ignored~\cite{Carena:2001fw}.  The label `4'  refers to
the would-be Goldstone boson of the $Z$ boson.  The absorptive part of
the Higgs-boson propagator matrix receives contributions from loops of
fermions, vector bosons, associated  pairs of Higgs and vector bosons,
Higgs-boson pairs, and sfermions:
\begin{equation}
\imag\widehat{\Pi}_{ij}(\hat{s})=
\imag\widehat{\Pi}^{ff}_{ij}(\hat{s})+
\imag\widehat{\Pi}^{VV}_{ij}(\hat{s})+\imag\widehat{\Pi}^{HV}_{ij}(\hat{s}) +
\imag\widehat{\Pi}^{HH}_{ij}(\hat{s}) +
\imag\widehat{\Pi}^{\tilde{f}\tilde{f}}_{ij}(\hat{s})\,,
\end{equation}
respectively. We refer  to Ref.~\cite{Ellis:2004fs} for their explicit
expressions.        For       the       Goldstone-Higgs       mixings,
$\imag\widehat{\Pi}_{i4\,,4i}$ and  $\imag\widehat{\Pi}_{44}$, we take
the leading contributions  ignoring all gauge-coupling mediated parts.
We  also  include the  $2\times2$-dimensional  propagator matrix  for  the
charged  Higgs  bosons   $D^{H^\pm}(\hat{s})$  spanned  by  the  basis
$(H^\pm\,,G^\pm)$, including off-diagonal absorptive parts:
\begin{equation}
D^{H^\pm}(\hat{s})=
\hat{s}\,\left(\begin{array}{cc}
\hat{s}-M_{H^\pm}^2+i\imag\widehat{\Pi}_{H^\pm H^\pm}(\hat{s}) &
i\imag\widehat{\Pi}_{H^\pm G^\pm}(\hat{s}) \\
i\imag\widehat{\Pi}_{G^\pm H^\pm}(\hat{s}) &
\hat{s}+i\imag\widehat{\Pi}_{G^\pm G^\pm}(\hat{s})
\end{array}
\right)^{-1}\,.
\end{equation}
The    relevant    Goldstone-boson     couplings    are    given    in
Appendix~\ref{sec:Goldstone}.
For  the 16  elements  of the  neutral  Higgs-boson propagator  matrix
$D^{H^0}({\hat{s}})$ and for the 4 elements of the charged Higgs-boson
propagator    matrix    $D^{H^\pm}({\hat{s}})$,    the   slots    {\tt
CAUX\_H(100-119)} are used as shown in Table~\ref{tab:caux}.
In Fig.~\ref{fig:dh3},  as an example,  we show the absolute  value of
all    components    of     the    Higgs-boson    propagator    matrix
$D^{H^0}({\hat{s}})$   as  functions   of  $\sqrt{\hat{s}}$   for  the
trimixing scenario with $\Phi_A=-\Phi_3=90^\circ$.

It   is  important  to   remark  that   the  $4\times   4$  propagator
matrix~(\ref{eq:hprop}) is sufficient to  encode all $H_i  - Z$- and $G^0 -
Z$ mixing     effects    within     the    Pinch     Technique    (PT)
framework~\cite{APNPB,PT}, which has been adopted here to remove consistently
gauge-dependent and  high-energy unitarity-violating terms from
$\imag    \widehat{\Pi}_{ij}    (\hat{s})$~\cite{Ellis:2004fs}.    For
example,    the    self-energy    transition    $H_i    \to    Z_\mu$,
$\widehat{\Pi}^\mu_{Z H_i} =  p^\mu \widehat{\Pi}_{Z H_i}$, is related
to $\widehat{\Pi}_{G^0 H_i}$ through
\begin{equation}
\hat{s}\; \widehat{\Pi}_{Z H_i} (\hat{s})\ =\ -\, i\, M^2_Z\; 
\widehat{\Pi}_{G^0  H_i} (\hat{s})\; ,  
\end{equation}
with $\hat{s}  = p^2$.  We recall that the  self-energy transitions $H_i\to
\gamma$  and  $G^0\to \gamma$  are  completely  absent  within the  PT
framework. More details may be found in~\cite{APNPB}.

Note  that  the  elements  of  the propagator  matrix  depend  on  the
center-of-mass energy,  denoted by $\sqrt{\hat{s}}$, which  is stored in
{\tt RAUX\_H(101)}, see Table~\ref{tab:raux}.
Along   with   $D^{H^0\,,H^\pm}(\hat{s})$,   the   $\hat{s}$-dependent
couplings   of    the   neutral   Higgs   bosons    to   two   gluons,
$S^{g}_i(\sqrt{\hat{s}})$   and  $P^{g}_i(\sqrt{\hat{s}})$,   and  two
photons,               $S^{\gamma}_i(\sqrt{\hat{s}})$              and
$P^{\gamma}_i(\sqrt{\hat{s}})$,  are  needed   when  we  consider  the
production of the neutral  Higgs bosons and study its CP properties at the 
LHC~\cite{gluon_fusion,lhc_cp,Ellis:2004fs}
and  a  $\gamma  \gamma$  collider~\cite{photon_collider,photon_cp,Ellis:2004hw}.   
They  are calculated   and   stored   in   {\tt   CAUX\_H(130-135)}   and   {\tt
CAUX\_H(140-145)} as shown in Table~\ref{tab:caux}.
We have included the dominant contributions coming from the 
$\tan\beta$ enhanced loops of sbottoms and gluinos and the subdominant ones 
coming from the stop-higgsino mediated diagrams.
Also included are
the resummed corrections to Yukawa couplings. 
For the electroweak corrections, see next subsection.
For the next-to-leading-order QCD corrections, appropriately calculated
$K$ factors should be taken into account separately in the calculation of
production cross sections~\cite{QCD1,QCD2}.

Two  additional  flags  are  used  to control  the  inclusion  of  the
off-diagonal    absorptive    parts   and    print    out   the    the
$\hat{s}$-dependent  propagator  matrix  and  the  $\hat{s}$-dependent
Higgs couplings to two photons and gluons:
\begin{itemize}
\item  {\tt  IFLAG\_H(13)}$=1$:  Does  not  include  the  off-diagonal
absorptive       parts      in      the       propagator      matrices
$D^{H^0\,,H^\pm}({\hat{s}})$.
\item  {\tt  IFLAG\_H(14)}$=1$:  Prints  out  each  component  of  the
Higgs-boson  propagator matrices $D^{H^0\,,H^\pm}({\hat{s}})$  and the
$\hat{s}$-dependent  couplings  $S^{\gamma\,,g}_i(\sqrt{\hat{s}})$ and
$P^{\gamma\,,g}_i(\sqrt{\hat{s}})$.
\end{itemize}

\subsection{Improved Treatment of Higgs-Boson Couplings and Decays}

The main updates include:
\begin{itemize}
\item The electroweak corrections to the neutral Higgs couplings to pairs of tau
leptons and $b$-quarks~\cite{Guasch:2001wv}. The explicit formulae used in the code
for the corrections, including non-vanishing CP phase effects, 
could be found in Ref.~\cite{Ellis:2004fs} and
Eqs.(A.1)-(A.2) of Ref.~\cite{cpsuperh}.
\item The three-body decay $H^+ \rightarrow t^* \bar{b} \rightarrow W^+ b \bar{b}$.
Some three-body decays play important role in Higgs searches~\cite{three_body}.
In addition to the three-body decays involving more than one massive gauge boson
considered previously, we include
the three-body decay $H^+ \rightarrow t^* \bar{b} \rightarrow W^+ b \bar{b}$ in the new
version.  The decay width is given by
\begin{eqnarray}
&&\hspace{-1.0cm}\Gamma(H^+ \rightarrow W^+ b \bar{b}) = \nonumber \\
&&\hspace{1.0cm}N_C\frac{g^2\,g_{tb}^2\,M_{H^\pm}}{512\pi^3}
\int_0^{1-\kappa_W}{\rm d}x_1
\int_{1-\kappa_W-x_1}^{1-\frac{\kappa_W}{1-x_1}}{\rm d}x_2
\frac{F(x_1,x_2)}{(1-x_2-\kappa_t+\kappa_b)^2+\kappa_t\gamma_t} ,
\end{eqnarray}
where $\kappa_x\equiv m_x^2/M_{H^\pm}^2$,
$\gamma_t\equiv \Gamma_t^2/M_{H^\pm}^2$ and $x_i\equiv 2E_i/M_{H^\pm}$
with $E_1$ and $E_2$ being the energies of the $b$ and $\bar{b}$ quarks, respectively.
In the charged Higgs-boson rest frame, the function $F(x_1,x_2)$ is given by
\begin{eqnarray}
F(x_1,x_2) &=&
\Bigg\{|g_L|^2\Bigg[\kappa_t\Bigg(\frac{(1-x_1)(1-x_2)}{\kappa_W}
+2x_1+2x_2-3+2\kappa_W\Bigg) -2\kappa_b\kappa_t\Bigg]
\nonumber \\ && 
+|g_R|^2\Bigg[\frac{x_2^3+x_1x_2^2-3x_2^2-2x_1x_2+3x_2+x_1-1}{\kappa_W}
\nonumber \\ && \hspace{1.5cm}
+(x_2^2+2x_1x_2-4x_2-2x_1+3-2\kappa_W)
\nonumber \\ && \hspace{1.5cm}
+\kappa_b\Bigg(-2x_1+3+2\kappa_W+
\frac{-2x_2^2-x_1x_2+5x_2+x_1-3}{\kappa_W}\Bigg)-2\kappa_b^2 \Bigg]
\nonumber \\ && 
+2\sqrt{\kappa_b\kappa_t}\,\real({g_Lg_R^*})\Bigg[
\frac{(x_2-1)^2}{\kappa_W}+(-x_2+1-2\kappa_W)+2\kappa_b\Bigg]\Bigg\} ,
\end{eqnarray}
where $g_L\equiv g^S_{H^+\bar{t}b}-ig^P_{H^+\bar{t}b}$ and
$g_R\equiv g^S_{H^+\bar{t}b}+ig^P_{H^+\bar{t}b}$.
\item The contributions from tau-lepton and charm-quark loops to the couplings
$S^\gamma_i(M_{H_i})$ and $P^\gamma_i(M_{H_i})$.
\item A new flag {\tt IFLAG\_H(57)}$=1$: This is an error message that appears 
when one of the magnitudes of the complex input parameters is negative.
\end{itemize}
The {\tt CPsuperH} homepage has been continuously brought up to date 
after its first appearance to include the updates discussed in this subsection and
others not mentioned here. We refer to the file {\tt 0LIST\_V1} 
for a full list of updates to the original version which can be found in the {\tt CPsuperH}
homepage.

\section{Higgs-Mediated Two-Loop Electric Dipole Moments}

\begin{figure}[t]
\hspace{ 0.0cm}
\vspace{-1.2cm}
\centerline{\epsfig{figure=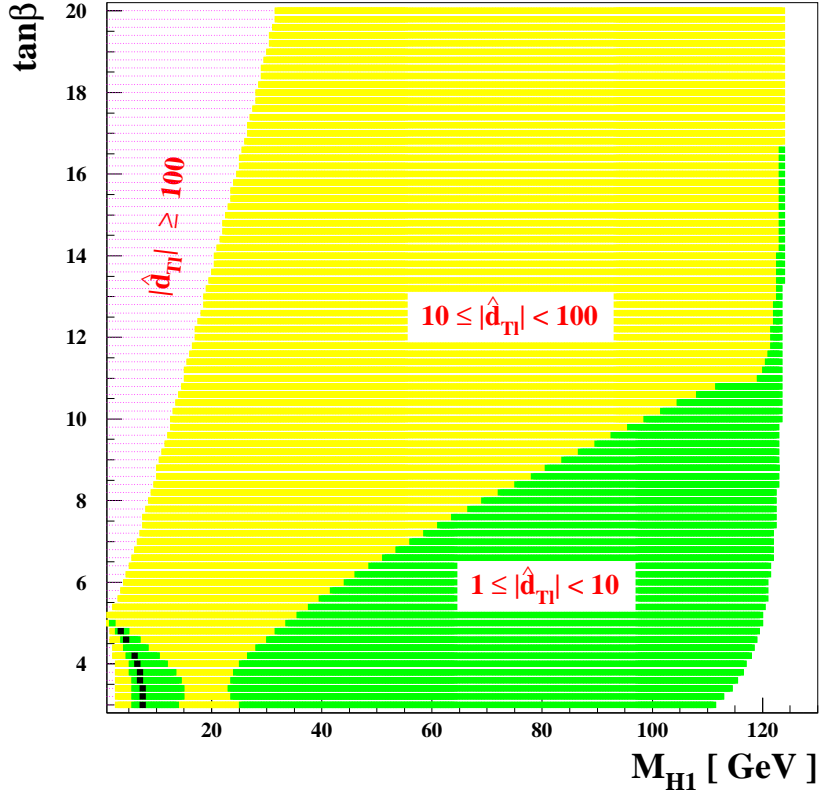,height=13.5cm,width=13.5cm}}
\vspace{-0.7cm}
\caption{\it
The Thallium EDM $\hat{d}_{\rm Tl} \equiv d^H_{\rm Tl}
\times 10^{24}$~$[e\,cm]$ in the CPX scenario with $\Phi_A=\Phi_3=90^\circ$
and $M_{\rm SUSY}=0.5$ TeV taking {\tt IFLAG\_H(12)}$=5$~\cite{Lee:2007ai}.
The different shaded regions correspond to different ranges of $|\hat{d}_{\rm Tl}|$ as
shown. Specially, in the narrow region denoted by black squares,
one has $|\hat{d}_{\rm Tl}|<1$,
consistent with the current Thallium EDM constraint.
}
\label{fig:dtl}
\end{figure}

The CP phases in the MSSM are significantly constrained by measurements 
of Electric Dipole Moments (EDMs).
In particular, the EDM of the Thallium atom may provide currently the most stringent 
constraint on MSSM scenarios with explicit CP violation.
The atomic EDM of $^{205}$Tl gets its main contributions
from two terms ~\cite{KL,PR}:
\begin{eqnarray}
d_{\rm Tl}\,[e\,cm]\ &=&\ -585\cdot d_e\,[e\,cm]\:
-\: 8.5\times 10^{-19}\,[e\,cm]\cdot (C_S\,{\rm TeV}^2)+ \cdots\,, \nonumber \\
&\equiv & (d_{\rm Tl})^e\,[e\,cm] + (d_{\rm Tl})^{C_S}\,[e\,cm] + \cdots\,,
\end{eqnarray}
where $d_e$ denotes the electron EDM and $C_S$ is the coefficient of
the CP-odd electron-nucleon interaction ${\cal
L}_{C_S}=C_S\,\bar{e}i\gamma_5 e\,\bar{N}N$. The dots denote
sub-dominant contributions from 6-dimensional tensor and
higher-dimensional operators.

The contributions of the first- and second-generation phases,
$\Phi_{A_{e,\mu}}$ and $\Phi_{A_{d,s}}$, to EDMs can be
drastically reduced either by assuming that these phases sufficiently
small, or if the first- and second-generation squarks and sleptons are
sufficiently heavy. However, even
when the contributions of the first and second generation phases to
EDMs are suppressed, there are still sizeable contributions to EDMs
from Higgs-mediated two-loop diagrams ~\cite{CKP}.

The Higgs-mediated two-loop Thallium ($d^H_{\rm Tl}$), electron ($d^H_e$), and 
muon ($d^H_\mu$) EDMs are calculated and stored in {\tt RAUX\_H(111-120)}
as shown in Table~\ref{tab:raux}. The Thallium and electron EDMs consist of:
\begin{eqnarray}
d^H_{\rm Tl}&=&(d^H_{\rm Tl})^e+(d^H_{\rm Tl})^{C_S}\,,\nonumber \\
d^H_e&=&(d^H_e)^{\tilde{t}}+(d^H_e)^{\tilde{b}}+(d^H_e)^t+(d^H_e)^b
+(d^H_e)^{\tilde{\chi}^\pm}\,.
\end{eqnarray}
The explicit expressions for the EDMs in the {\tt CPsuperH} conventions and notations may be
found in Ref.~\cite{Ellis:2005ik}. A flag {\tt IFLAG\_H(15)}$=1$ is used to print out the
results of the EDM calculations:
\begin{itemize}
\item {\tt IFLAG\_H(15)}$=1$: Print out EDMs.
\end{itemize}

In Fig.~\ref{fig:dtl}, we show the rescaled Thallium EDM $\hat{d}_{\rm
Tl} \equiv d^H_{\rm Tl} \times 10^{24}$ in units of $e\,cm$ in the
$\tan\beta$-$M_{H_1}$ plane, in the CPX scenario with {\tt IFLAG\_H(12)}$=5$.
We observe, when $\tan\beta \lsim 5$ and $M_{H_1}\lsim 10$ GeV,
one may have $|\hat{d}_{\rm Tl}|<1$ only in the narrow region denoted by black 
squares which
is consistent with the current 2-$\sigma$ upper bound on the Thallium EDM
\cite{Regan:2002ta}: $|d_{\rm Tl}|\ \lsim\ 1.3\times 10^{-24}\,[e\,cm]$.
We note that the region $8~{\rm GeV}\lsim M_{H_1}\lsim 10$ GeV 
with $\tan\beta \lsim 10$ 
has not been excluded by the combined constraints from 
the LEP searches~\cite{LEP_HIGGS}
and the $\Upsilon(1S)\rightarrow \gamma H_1$ decay~\cite{upsilon_visible}.

The Thallium EDM constraint can be evaded by assuming cancellations 
between the two-loop contributions considered here and possible
one-loop contributions which depend  on
different CP-odd  phases related  to  the first and second generations
of squarks and sleptons. For example, assuming cancellation of less than 1 part
in 10, the region with $1 \leq |\hat{d}_{\rm Tl}| < 10$ in Fig.~\ref{fig:dtl} is
allowed.
In the future, this treatment of the most important two-loop
contributions to the Thallium EDM will be supplemented by a more
complete implementation of calculations of the well-known 1-loop
contributions to this and other EDMs.

\section{$B$-Meson Observables}

An important innovation in {\tt CPsuperH2.0} is the inclusion of
the following important Higgs-mediated 
$B$-meson observables:
\begin{itemize}
\item The branching ratio of $B_s$ meson into a pair of muons: $B(B_s\to \mu\mu)$,
\item The branching ratio of $B_d$ meson into a pair of tau leptons: $B(B_d\to \tau\tau)$,
\item The SUSY contribution to the $B_d^0$-$\bar{B}_d^0$ mass difference: 
$\Delta M_{B_d}^{\rm SUSY}$,
\item The SUSY contribution to the $B_s^0$-$\bar{B}_s^0$ mass difference: 
$\Delta M_{B_s}^{\rm SUSY}$,
\item The ratio of the branching ratio $B(B_u\to \tau\nu)$ to the SM value:\\
$$R_{B\tau\nu}=\frac{B(B_u^-\to \tau\nu)}{B^{\rm SM}(B_u^-\to \tau\nu)}$$,
\item The branching ratio $B(B\to X_s \gamma)$ and the direct CP asymmetry ${\cal A}_{\rm
CP}(B\to X_s\gamma)$.
\end{itemize}
We adopt the most recent gauge-invariant and flavour-covariant formalism to calculate
the flavour-changing effective Lagrangian for the interactions of the
neutral and charged Higgs fields to the up- and down-type quarks including a new class of
dominant subleading contributions~\cite{MCPMFV}.
In the current version, the single-Higgs insertion approximation is used.

For the calculations of $B$-meson observables, the array {\tt SMPARA\_H} for the SM
parameters has been
extended to include information on the CKM matrix,
parameterized via $\lambda$, $A$, $\bar\rho$, and
$\bar\eta$, as seen in Table~\ref{tab:smpara}. The CKM matrix is constructed as~\cite{PDG}
\begin{equation}
V=\left(
\begin{array}{ccc}
c_{12}\,c_{13} & s_{12}\,c_{13} & s_{13}\,e^{-i\,\delta} \\
-s_{12}\,c_{23}-c_{12}\,s_{23}\,s_{13}\,e^{i\,\delta} & 
c_{12}\,c_{23}-s_{12}\,s_{23}\,s_{13}\,e^{i\,\delta} & s_{23}\,c_{13} \\
s_{12}\,s_{23}-c_{12}\,c_{23}\,s_{13}\,e^{i\,\delta} & 
-c_{12}\,s_{23}-s_{12}\,c_{23}\,s_{13}\,e^{i\,\delta} & c_{23}\,c_{13} 
\end{array}
\right)\,,
\end{equation}
where $s_{ij}=\sin\theta_{ij}$, $c_{ij}=\cos\theta_{ij}$, and $\delta$ is the KM phase
with $s_{ij}\,,c_{ij} \geq 0$. In terms of $\lambda$, $A$, $\bar\rho$, and
$\bar\eta$, they are given by
\begin{equation}
s_{12}=\lambda \,, \ \ \ s_{23}=A\lambda^2\,, \ \ \
s_{13}\,e^{i\,\delta}=\frac{A\lambda^3(\bar\rho+i\,\bar\eta)\sqrt{1-A^2\lambda^4}}
{\sqrt{1-\lambda^2}\left[1-A^2\lambda^4(\bar\rho+i\,\bar\eta)\right]}\,,
\end{equation}
and $c_{ij}=\sqrt{1-|s_{ij}|^2}$.
The SUSY parameter array {\tt SSPARA\_H} is also extended to include the 
hierarchy factors
$\rho_{\tilde{Q},\tilde{U},\tilde{D},\tilde{L},\tilde{E}}$ between the first two and 
third generations~\cite{DP}, see Table~\ref{tab:sspara}. In the super-CKM basis, 
the $3\times 3$ squark mass matrices squared are taken to be diagonal:
\begin{eqnarray}
{\bf \widetilde{M}}^2_Q \ &=& \ m_{\tilde{Q}_3}^2 \
\times \ {\rm diag}\,(\rho_{\tilde{Q}}^2,\rho_{\tilde{Q}}^2,1)\,, \nonumber \\
{\bf \widetilde{M}}^2_U \ &=& \ m_{\tilde{U}_3}^2 \
\times \ {\rm diag}\,(\rho_{\tilde{U}}^2,\rho_{\tilde{U}}^2,\,1)\,, \nonumber \\
{\bf \widetilde{M}}^2_D \ &=& \ m_{\tilde{D}_3}^2 \
\times \ {\rm diag}\,(\rho_{\tilde{D}}^2,\rho_{\tilde{D}}^2,1)\,, \nonumber \\
{\bf \widetilde{M}}^2_L \ &=& \ m_{\tilde{L}_3}^2 \
\times \ {\rm diag}\,(\rho_{\tilde{L}}^2,\rho_{\tilde{L}}^2,1)\,, \nonumber \\
{\bf \widetilde{M}}^2_E \ &=& \ m_{\tilde{E}_3}^2 \
\times \ {\rm diag}\,(\rho_{\tilde{E}}^2,\rho_{\tilde{E}}^2,1)\,.
\end{eqnarray}

Finally, the results for the $B$-meson observables are stored in {\tt RAUX\_H(130-136)} as 
shown in Table~\ref{tab:raux}. The SUSY  contributions to the  $\Delta B=2$ transition
amplitudes are stored in {\tt CAUX\_H(150)} and {\tt CAUX\_H(151)}, see Table~\ref{tab:raux}.
Note the relations ${\tt RAUX\_H(132)}=2\times|{\tt CAUX\_H(150)}|$ and
${\tt RAUX\_H(133)}=2\times|{\tt CAUX\_H(151)}|$.
Two flags {\tt IFLAG\_H(16)} and {\tt IFLAG\_H(17)}
are used to print out the results of the calculation of
$B$-meson observables:
\begin{itemize}
\item {\tt IFLAG\_H(16)}$=1$: Print out $B$-meson observables.
\item {\tt IFLAG\_H(17)}$=1$: Print out details of the $B\to X_s \gamma$ calculation.
\end{itemize}

For numerical examples of $B$-meson observables, we take the CPX 
scenario~\cite{Carena:2000ks} with $M_{\rm SUSY}=0.5$ TeV and the 
common $A$-term phase $\Phi_A\equiv\Phi_{A_t}=\Phi_{A_t}=\Phi_{A_\tau}$ 
in the convention $\Phi_\mu=0^\circ$. 
We take account of the dependence on the hierarchy factors
$\rho_{\tilde{Q},\tilde{U},\tilde{D}}$ between the first two and the third generations,
taking a common value $\rho$ for the three of them.

\begin{figure}[htb]
\hspace{ 0.0cm}
\vspace{-0.5cm}
\centerline{\epsfig{figure=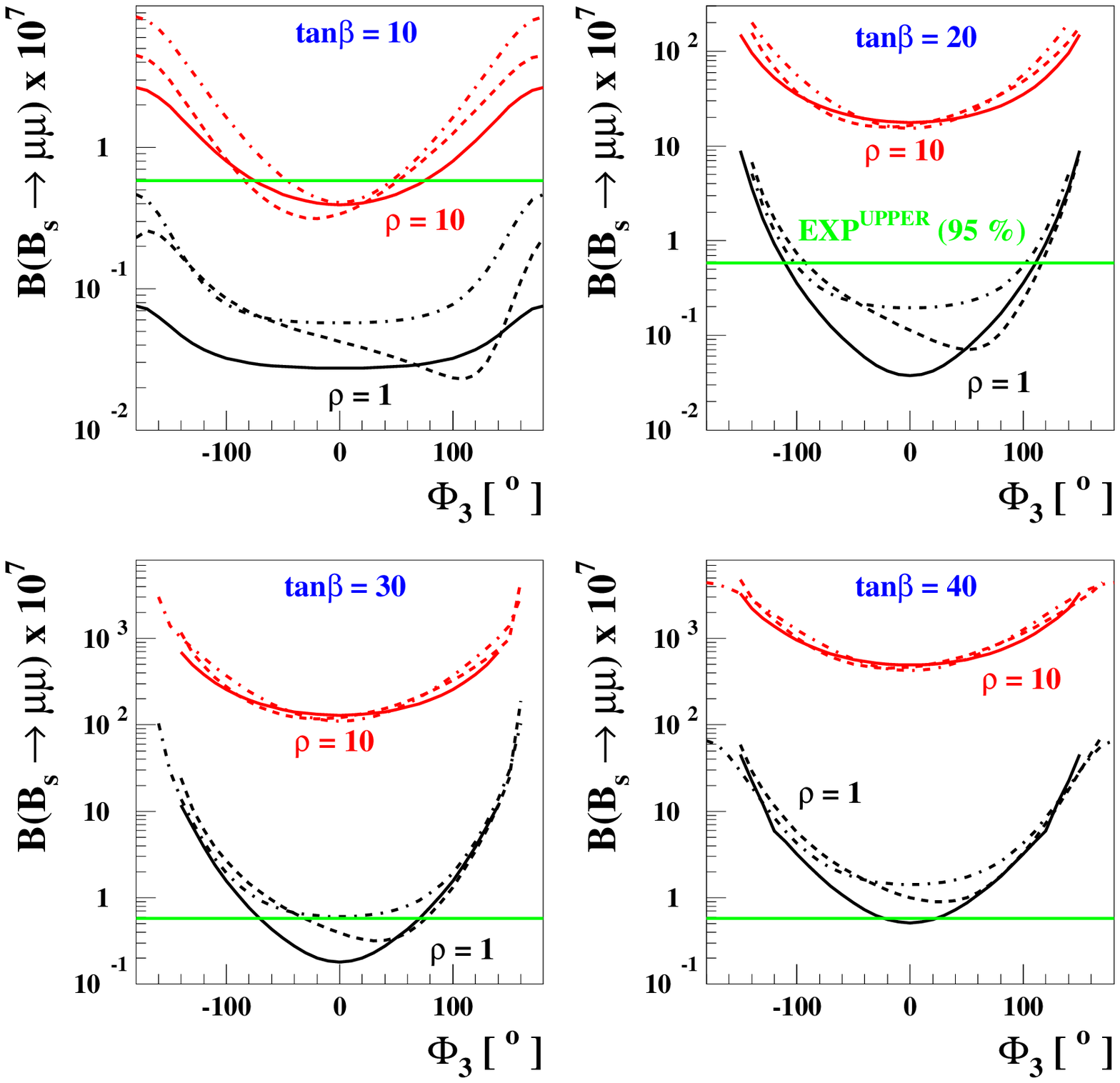,height=14cm,width=14cm}}
\vspace{-0.5cm}
\caption{\it The branching ratio $B(B_s \to \mu^+\mu^-) \times 10^7$ as a function of
$\Phi_3$ for four values of $\tan\beta$: $\tan\beta=10$ (upper left), 20 (upper right), 
30 (lower left), and 40 (lower right).
The CPX scenario is taken with $M_{\rm SUSY}=0.5$ TeV and $M_{H^\pm}=200$ GeV
in the convention $\Phi_\mu=0$. In each frame, the lower three lines are for the case
$\rho\equiv\rho_{\tilde{Q}}=\rho_{\tilde{U}}=\rho_{\tilde{D}}=1$ and the upper lines for 
$\rho=10$ where the solid, dashed, and dash-dotted lines are for $\Phi_A=0^\circ$, 
$90^\circ$, and $180^\circ$, respectively. The current 95 \% experimental upper 
bound, $5.8 \times 10^{-8}$~\cite{CDF:2007kv}, is also shown as
a horizontal line in each frame.  }
\label{fig:bsmm_p3}
\end{figure}

\begin{figure}[htb]
\hspace{ 0.0cm}
\vspace{-0.5cm}
\centerline{\epsfig{figure=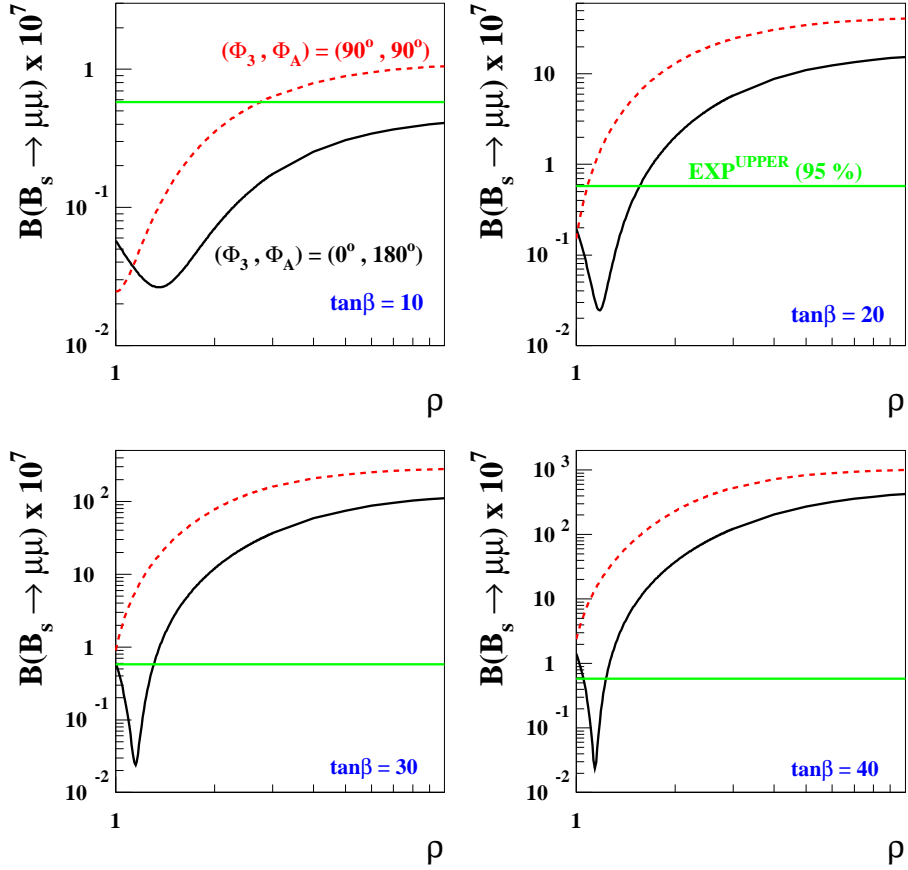,height=14cm,width=14cm}}
\vspace{-0.5cm}
\caption{\it The branching ratio $B(B_s \to \mu^+\mu^-) \times 10^7$ as a function of the
common hierarchy factor $\rho\equiv\rho_{\tilde{Q}}=\rho_{\tilde{U}}=\rho_{\tilde{D}}$
for four values of $\tan\beta$: $\tan\beta=10$ (upper left), 20 (upper right), 30,
(lower left), and 40 (lower right).
The CPX scenario is taken with $M_{\rm SUSY}=0.5$ TeV and $M_{H^\pm}=200$ GeV
in the convention $\Phi_\mu=0$. In each frame, the solid line is for 
$(\Phi_3\,,\Phi_A)=(0^\circ\,,180^\circ)$ and the dashed one for $(90^\circ\,,90^\circ)$.
The current 95 \% experimental upper bound, $5.8 \times
10^{-8}$~\cite{CDF:2007kv}, is also shown as
a horizontal line in each frame.  }
\label{fig:bsmm_rho}
\end{figure}

\begin{figure}[htb]
\hspace{ 0.0cm}
\vspace{-0.5cm}
\begin{center}
\includegraphics[width=7.5cm]{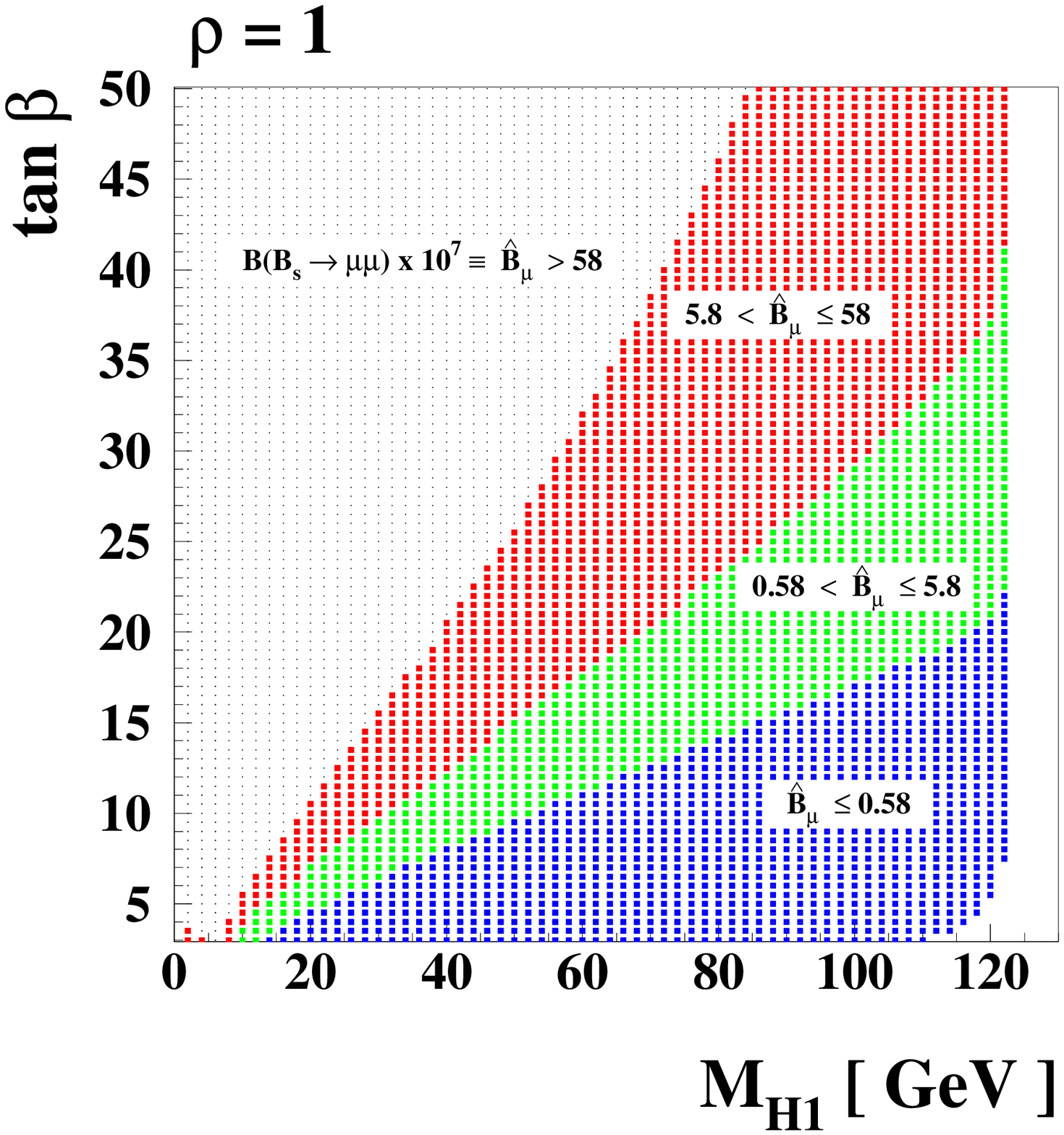}
\includegraphics[width=7.5cm]{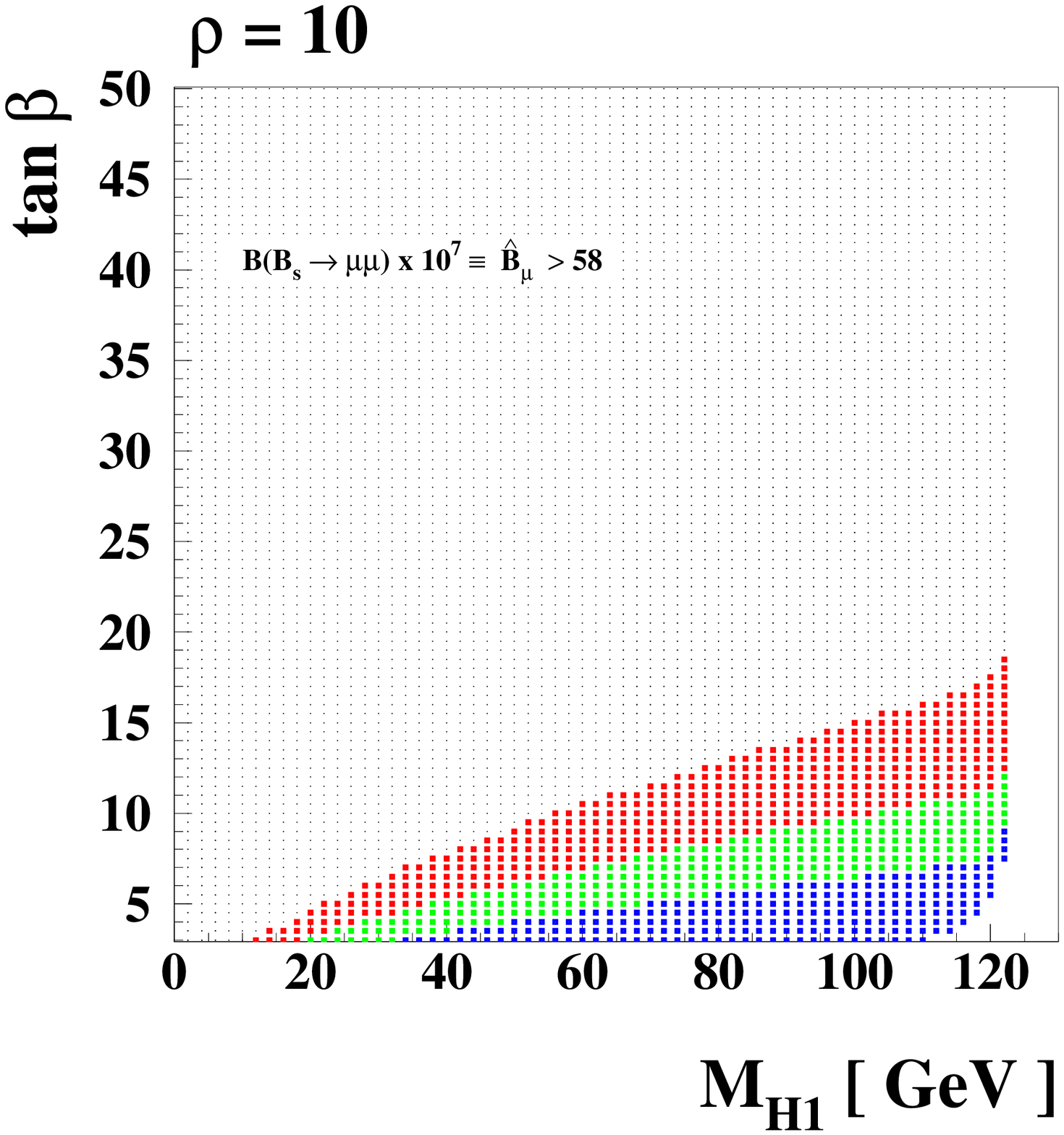}
\end{center}
\vspace{-0.5cm}
\caption{\it The branching ratio $\widehat{B}_\mu\equiv B(B_s \to \mu^+\mu^-) \times 10^7$
in the $(\tan\beta\,,M_{H_1})$ plane.  The CPX scenario is taken
with $\Phi_A=\Phi_3=90^\circ$ and $M_{\rm SUSY}=0.5$ TeV for two values of
the common hierarchy factor: $\rho=1$ (left) and $10$ (right). 
The unshaded region is not theoretically allowed.
The different shaded regions correspond to different ranges of
$\widehat{B}_\mu$, as shown: specifically, 
$\widehat{B}_\mu < 0.58$ in the lowest (blue) low-$\tan\beta$ 
region, consistent with the current
upper limit at 95 \% C.L.  }
\label{fig:bsmm_mh1tb}
\end{figure}

Figure~\ref{fig:bsmm_p3} shows the dependence of the branching ratio
$B(B_s\to\mu^+\mu^-)$ on the phase of the
gluino mass parameter $\Phi_3$ for four values of $\tan\beta$. The charged Higgs-boson
pole mass is fixed at $M_{H^\pm}=200$ GeV. In each frame, two sets of three lines
are shown. The upper lines are for higher $\rho=10$ and the lower ones for $\rho=1$. 
For fixed $\rho$, three lines show the cases of $\Phi_A=0^\circ$ (solid), 
$90^\circ$ (dashed), and $180^\circ$ (dash-dotted). The $\rho$ dependence is shown in
Fig.~\ref{fig:bsmm_rho}. We clearly see the {\it GIM operative point} mechanism
discussed in Ref.~\cite{DP} around $\rho\sim 1.2$ when
$(\Phi_3\,,\Phi_A)=(0^\circ\,,180^\circ)$ (solid lines).  
Figure~\ref{fig:bsmm_mh1tb} shows
the rescaled branching ratio $\widehat{B}_\mu\equiv B(B_s\to\mu^+\mu^-)\times 10^7$ 
in the $M_{H_1}$-$\tan\beta$ plane
when the phases are fixed at $\Phi_A=\Phi_3=90^\circ$. The unshaded region is not
theoretically allowed. Only the region with $\widehat{B}_\mu < 0.58$ is consistent
with the current experimental upper limit at 95 \% C.L., corresponding to
$\tan\beta \lsim 20\,(8)$ for $\rho=1\,(10)$.

\begin{figure}[htb]
\hspace{ 0.0cm}
\vspace{-0.5cm}
\begin{center}
\includegraphics[width=7.5cm]{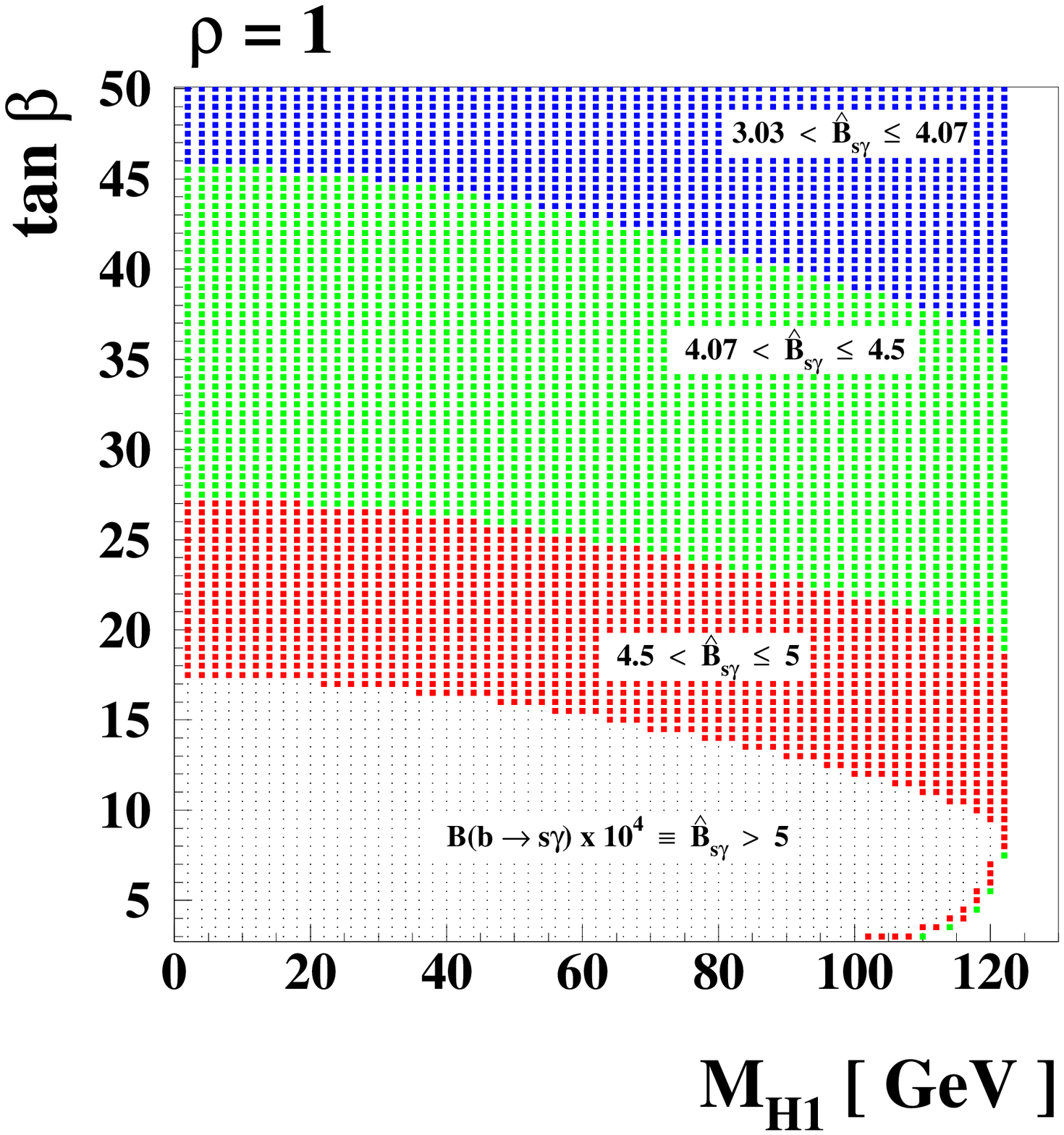}
\includegraphics[width=7.5cm]{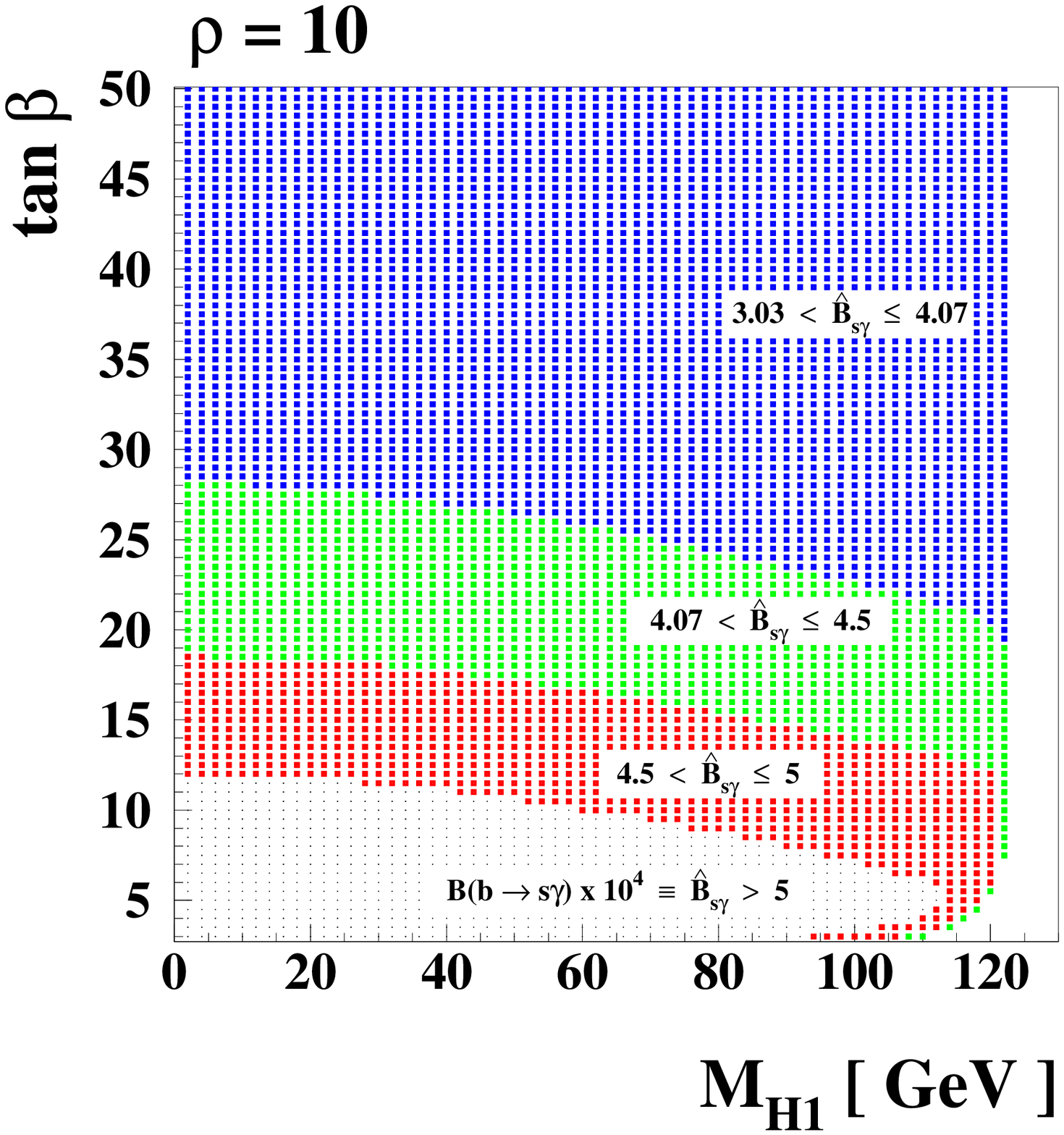}
\end{center}
\vspace{-0.5cm}
\caption{\it The branching ratio $\widehat{B}_{s\gamma}\equiv B(B \to X_s \gamma) 
\times 10^4$ in the $(\tan\beta\,,M_{H_1})$ plane.  The same CPX scenario with
$\Phi_A=\Phi_3=90^\circ$ is taken 
as in Fig.~\ref{fig:bsmm_mh1tb}.
The different shaded regions correspond to different ranges of
$\widehat{B}_{s\gamma}$, as shown: specifically, $3.03<\widehat{B}_{s\gamma} \leq 4.07$ 
in the upmost (blue) high-$tan\beta$ region,
consistent with the current experimentally allowed 
2-$\sigma$ region, $3.03<\widehat{B}_{s\gamma} \leq 4.07$~\cite{HFAG}.}
\label{fig:bsg_mh1tb}
\end{figure}

The rescaled branching ratio 
$\widehat{B}_{s\gamma}\equiv B(B\to X_s\gamma)\times 10^4$
is shown in Fig.~\ref{fig:bsg_mh1tb}. In contrast to the $B_s\to\mu^+\mu^-$ case, 
we observe that higher $\tan\beta$ region is experimentally allowed: 
$\tan\beta \gsim 35\,(20)$ for $\rho=1\,(10)$. This is because the charged-Higgs
contribution is suppressed due to the threshold corrections when $\tan\beta$ is large.
The charged-Higgs contribution to $B\to X_s\gamma$ is proportional 
to $1/(1+|\kappa|^2\tan^2\beta)$~\cite{Carena:2000uj}, 
where $\kappa$ represents the threshold corrections with
$|\kappa|\simeq 0.05$ for the parameters chosen~\cite{Borzumati:2004rd}.

\begin{figure}[htb]
\hspace{ 0.0cm}
\vspace{-0.5cm}
\begin{center}
\includegraphics[width=7.5cm]{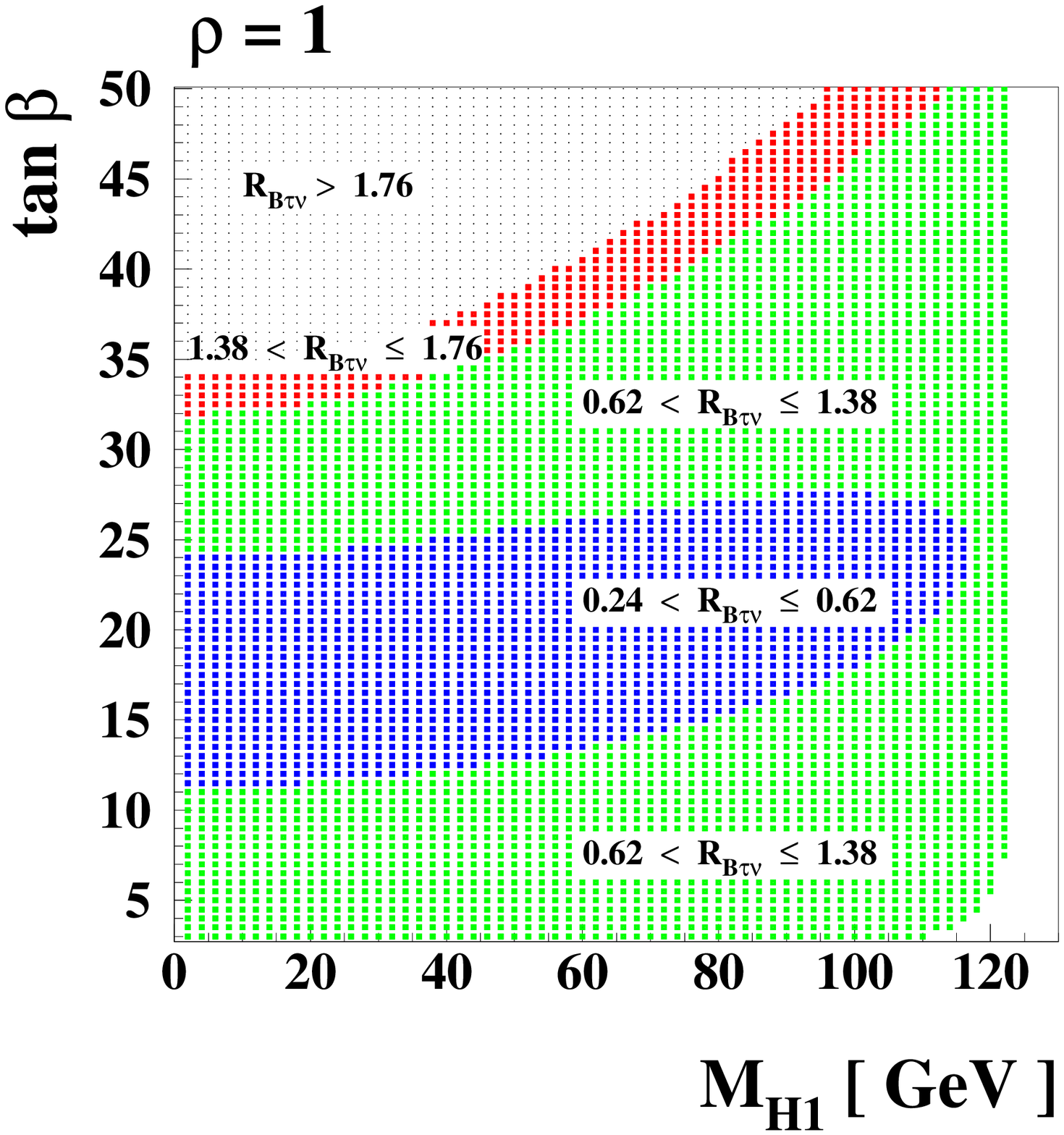}
\includegraphics[width=7.5cm]{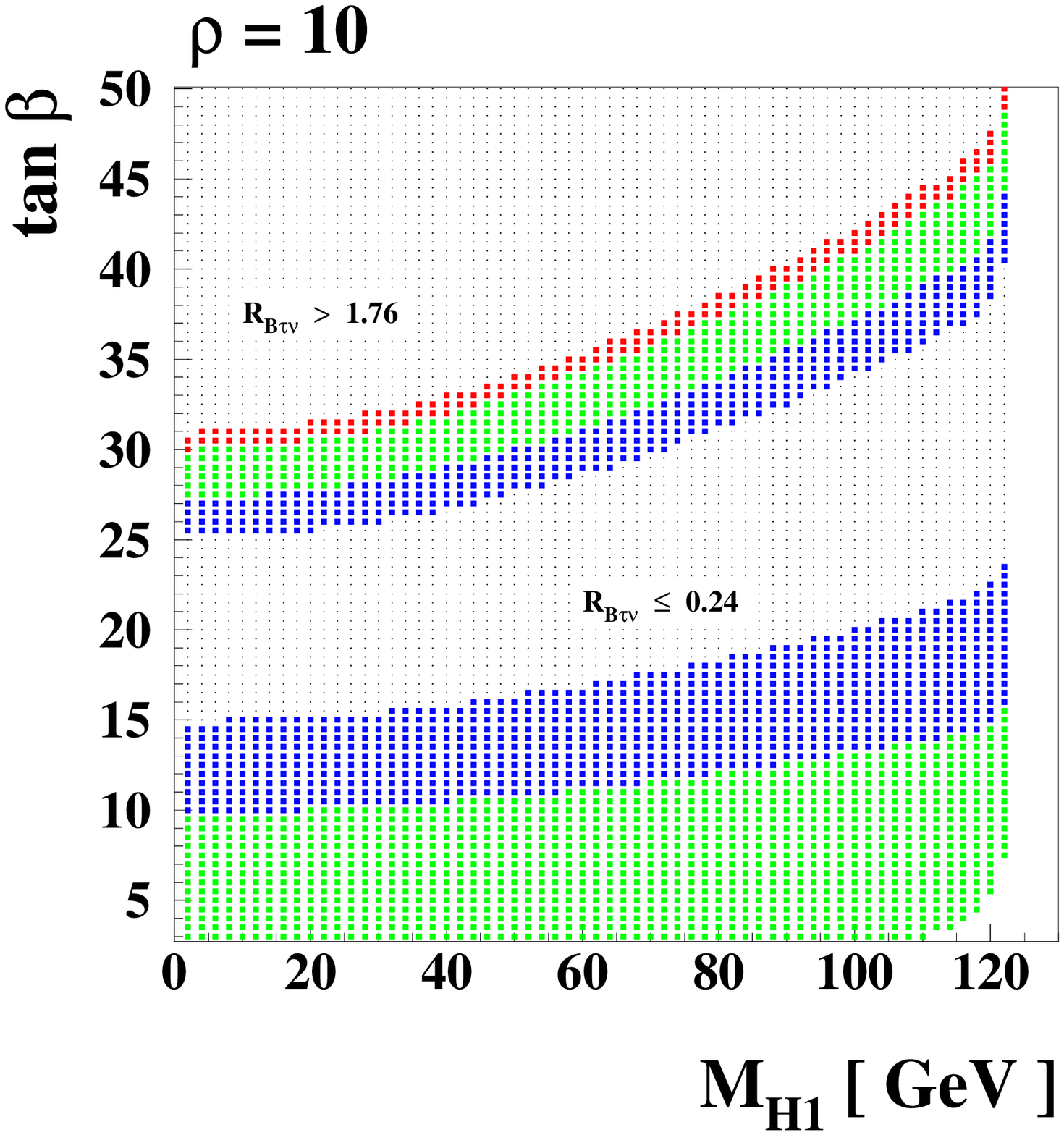}
\end{center}
\vspace{-0.5cm}
\caption{\it The ratio $R_{B\tau\nu}$ in the $(\tan\beta\,,M_{H_1})$ plane.  
The same CPX scenario with $\Phi_A=\Phi_3=90^\circ$ 
is taken as in Fig.~\ref{fig:bsmm_mh1tb} for two values of $\rho$:
$\rho=1$ (left) and $10$ (right).
The different shaded regions correspond to the regions allowed at the
1-$\sigma$ and 2-$\sigma$ levels
by the recent BELLE and BABAR results: $R_{B\tau\nu}^{\rm
EXP}=1.0\pm 0.38$~\cite{Btaunu,MCPMFV}. In the right frame, specifically,
the 2-$\sigma$ excluded
regions are shown as $R_{B\tau\nu} > 1.76$ (in the high-$\tan\beta$ region)  
and $R_{B\tau\nu} \leq 0.24$ (in the middle-$\tan\beta$ region). }
\label{fig:rbtn_mh1tb}
\end{figure}

Figure~\ref{fig:rbtn_mh1tb} shows the ratio of the branching ratio $B(B_u\to \tau\nu)$ to 
its SM value, $R_{B\tau\nu}$. In the left frame with $\rho=1$, we see
two connected bands of the experimentally  
allowed 1-$\sigma$  region, $0.62 < R_{B\tau\nu} <1.38$. If we consider the
2-$\sigma$ limit, only the upper-left region with
$M_{H_1}\lsim 95$ GeV and $\tan\beta \gsim 35$ 
is not allowed. For larger  $\rho=10$, the
allowed region becomes narrower.
 
\begin{figure}[htb]
\hspace{ 0.0cm}
\vspace{-0.5cm}
\begin{center}
\includegraphics[width=7.5cm]{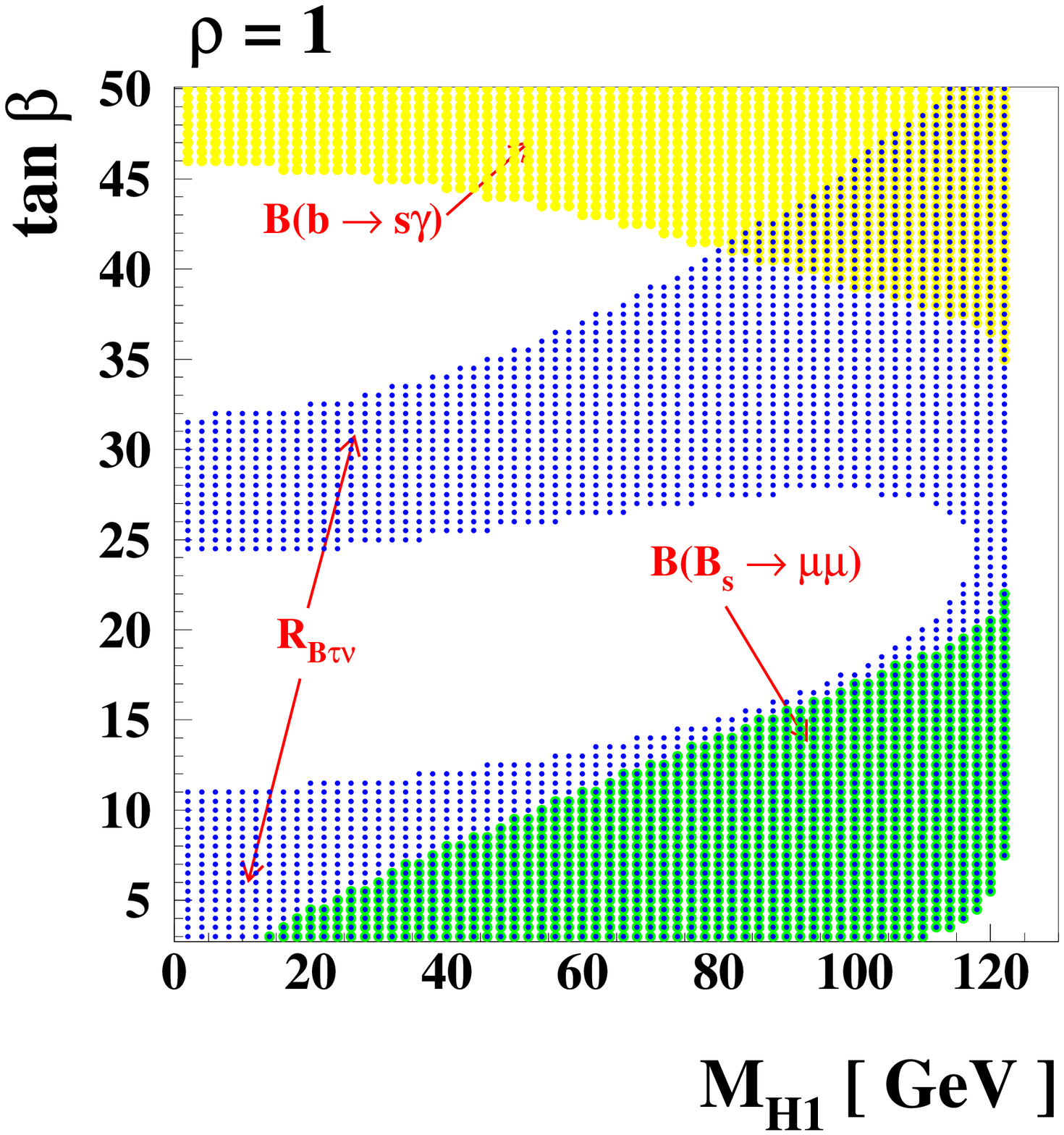}
\includegraphics[width=7.5cm]{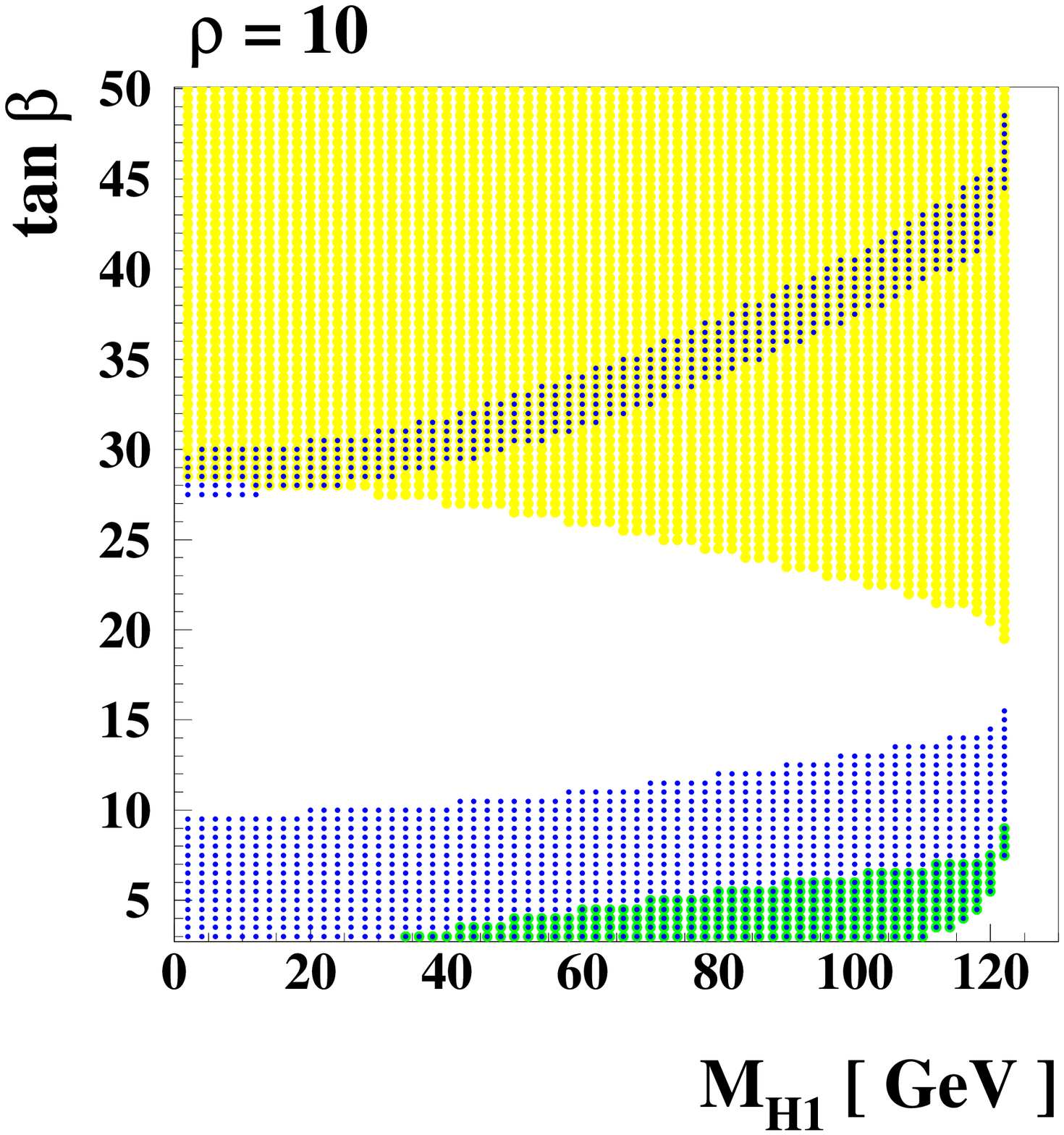}
\end{center}
\vspace{-0.5cm}
\caption{\it The experimental constraints from
$B(B_s \to \mu^+\mu^-)$ (95 \%), $B(B \to X_s\gamma)$ (2 $\sigma$), and $R_{B\tau\nu}$
(1 $\sigma$) in the $(\tan\beta\,,M_{H_1})$ plane for two values of $\rho$. The same CPX
scenario with $\Phi_A=\Phi_3=90^\circ$ is taken as in Fig.~\ref{fig:bsmm_mh1tb}.
}
\label{fig:three_mh1tb}
\end{figure}

In Fig.~\ref{fig:three_mh1tb}, we show the region satisfying the 
experimental constraints from $B(B_s \to \mu^+\mu^-)$ (95 \%), 
$B(B \to X_s\gamma)$ (2 $\sigma$), and $R_{B\tau\nu}$ (1 $\sigma$). First we observe
that there is no region that satisfies the $B_s \to \mu^+\mu^-$ and
$B \to X_s\gamma$ constraints simultaneously for both $\rho=1$ and $10$. 
If one neglects the constraint
from $B(B_s \to \mu^+\mu^-)$, only the high-$\tan\beta$ region would remain.
Taking account of $B_u\to \tau\nu$ constraint, the region with
$\tan\beta\gsim 36$ and $M_{H_1}\gsim 80$ GeV is allowed when $\rho=1$. 
On the other hand, neglecting the constraint from $B(B \to X_s\gamma)$, the allowed region is
constrained in the parameter space with $\tan\beta\lsim 20$ and $M_{H_1}\gsim 10$ GeV 
for $\rho=1$.  For $\rho=10$, the $B \to X_s\gamma$ constraint is relaxed but those
from $B(B_s \to \mu^+\mu^-)$ and $R_{B\tau\nu}$ become more stringent.

\begin{figure}[htb]
\hspace{ 0.0cm}
\vspace{-0.5cm}
\centerline{\epsfig{figure=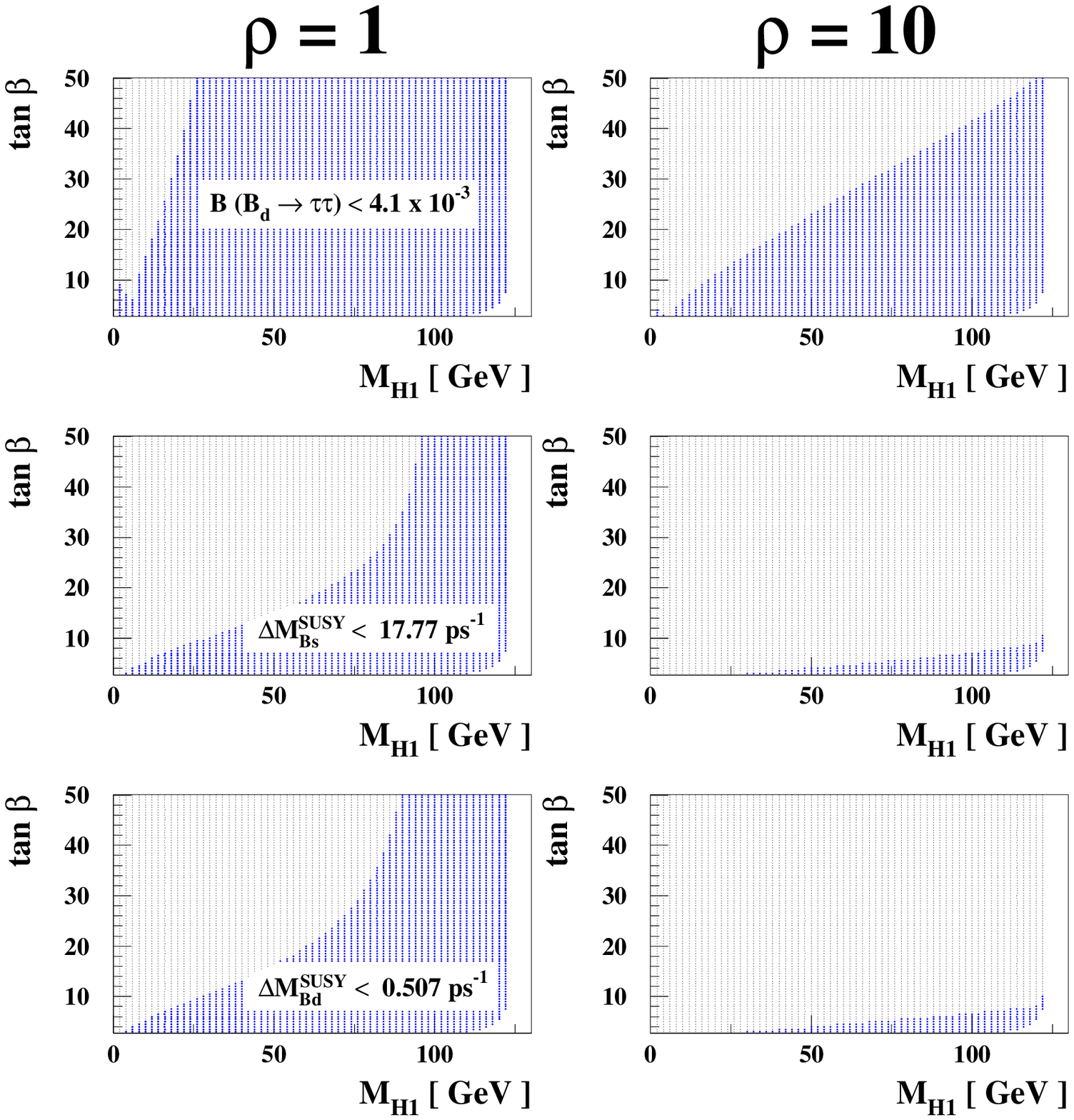,height=14cm,width=14cm}}
\vspace{-0.5cm}
\caption{\it The region allowed experimentally by the measurement $B(B_d\to
\tau^+\tau^-) < 4.1 \times 10^{-3}$ (90 \%)~\cite{BDTT} (upper frames) and the regions
where the SUSY contribution is smaller than the measured values of
$B_s^0$-$\bar{B}_s^0$ mass difference~\cite{Evans:2007hq} (middle frames)
and $B_d^0$-$\bar{B}_d^0$ mass difference~\cite{PDG} (lower frames), in the
$(\tan\beta\,,M_{H_1})$ plane.  The left three
frames are for $\rho=1$ and the right ones for $\rho=10$.  The same CPX
scenario with $\Phi_A=\Phi_3=90^\circ$ is taken as in Fig.~\ref{fig:bsmm_mh1tb}.  }
\label{fig:bdtt_dmb}
\end{figure}

Finally, in Fig.~\ref{fig:bdtt_dmb}, we show 
the region allowed experimentally by the measurement $B(B_d\to
\tau^+\tau^-) < 4.1 \times 10^{-3}$ (90 \%)~\cite{BDTT} (upper frames) and the regions
where the SUSY contribution is smaller than the measured values of
$B_s^0$-$\bar{B}_s^0$ mass difference~\cite{Evans:2007hq} (middle frames)
and $B_d^0$-$\bar{B}_d^0$ mass difference~\cite{PDG} (lower frames).
We see that the $B(B_d \to \tau^+\tau^-)$ constraint has the least impact on
these parameter planes, whereas the impacts of the $B_s^0$-$\bar{B}_s^0$
and $B_d^0$-$\bar{B}_d^0$ mass differences are similar.

These examples illustrate the possible interplays between the different
$B$-meson observables, and how they may vary significantly with the values of
the CP-violating phases. {\tt CPsuperH2.0} provides a unique tool for combining
these constraints and pursuing their implications for other observables.
In the future, the {\tt CPsuperH2.0} treatment of these important $B$-meson observables
 will be supplemented by the implementation of calculations
of other flavour observables, including the $K$ sector.

\section{Summary and Outlook}\label{sec:summary}

We have presented in this paper a description of the new features of
the  Fortran   code  {\tt  CPsuperH2.0}.   In   addition  to  improved
calculations  of  the  Higgs-boson  poles masses  with  more  complete
treatment of threshold effects  in self-energies and Yukawa couplings,
the  {\em  complete}  $4\times  4$  ($2\times  2$)  neutral  (charged)
Higgs-boson  propagator  matrices   with  the  Goldstone-Higgs  mixing
effects have been consistently implemented.  Specifically, the neutral
Higgs-boson propagator  matrix constitutes a  necessary ingredient for
the studies of a system  of strongly-mixed Higgs bosons at colliders
together  with the center-of-mass  dependent Higgs-boson  couplings to
gluons  and  photons.   It  also  provides  the  improved  Higgs-boson
couplings to tau  leptons, $b$ quarks, and two  photons.  The important
three-body decay $H^+ \to t^* \bar{b} \to W^+ b \bar{b}$ is included.

In order to provide a more complete, consistent tool for calculating
CP-violating observables in the MSSM, and specifically to incorporate 
the important constraints coming from precision experiments at
low energies, {\tt CPsuperH2.0} has  been extended to include a number
of  $B$-meson  observables, as  well  as  the Higgs-mediated  two-loop
contributions to EDMs of the Thallium atom, electron and muon.
The currently available $B$-meson observables are the branching ratios
of $B_s \to \mu^+ \mu^-$, $B_d \to \tau^+ \tau^-$, $B_u \to \tau \nu$,
$B  \to X_s  \gamma$ and  the latter's  CP-violating  asymmetry ${\cal
A}_{\rm CP}$, and the supersymmetric contributions to the $B^0_{s,d} -
{\bar  B^0_{s,d}}$ mass  differences.  Further  low-energy observables
are to be included in future updates.

The improved  Fortran code {\tt  CPsuperH2.0} provides a  coherent and
complete numerical  framework in which one  can calculate consistently
observables in  both low- and high-energy  experiments probing physics
beyond the SM.

\vspace{-0.2cm}
\subsection*{Acknowledgements}
\vspace{-0.3cm}
\noindent
The  work of  J.S.L.   was supported  in  part by  the Korea  Research
Foundation and the Korean Federation of  Science and Technology
Societies Grant  funded  by  the Korea  Government  (MOEHRD, Basic  Research
Promotion Fund) and in part by the National Science Council of Taiwan, R.O.C.
under Grant No. NSC 96-2811-M-008-068.
The work of  A.P. was support  in part by  the STFC
research grant: PP/D000157/1. 
Work at ANL is supported in part by the US DOE, Div. of HEP, Contract
DE-AC02-06CH11357 . Fermilab is operated by Universities Research Association  Inc.
under contract no. DE-AC02-76CH02000 with the DOE.
We thank S.Y.~Choi and M.~Drees for past
collaboration on {\tt CPsuperH}, and for discussions on this updated version.

\vspace{-0.5cm}


\vspace{-1.5cm}

\newpage


\def\theequation{\Alph{section}.\arabic{equation}}
\begin{appendix}

\setcounter{equation}{0}
\section{List of changes}
Here we summarize the improved features
introduced in {\tt CPsuperH2.0} compared to the prior version
of {\tt CPsuperH}.
\begin{itemize}
\item New common blocks:
\begin{itemize}
\item {\tt COMMON /HC\_RAUX/ RAUX\_H(NAUX$=$999)}, see Table~\ref{tab:raux}
\item {\tt COMMON /HC\_CAUX/ CAUX\_H(NAUX$=$999)}, see Table~\ref{tab:caux}
\end{itemize}
\item Extended arrays for input parameters:
\begin{itemize}
\item {\tt SMPARA\_H(NSMIN$=$19)}, see Table~\ref{tab:smpara}
\item {\tt SMPARA\_H(NSSIN$=$26)}, see Table~\ref{tab:sspara}
\end{itemize}
\item New names for improved {\tt FORTRAN} files:
\begin{itemize}
\item {\tt cpsuperh.f} $~\longrightarrow ~$ {\tt cpsuperh2.f}
\item {\tt fillpara.f} $~\longrightarrow ~$ {\tt fillpara2.f}
\item {\tt fillhiggs.f} $\,\longrightarrow ~$ {\tt fillhiggs2.f}
\item {\tt fillcoupl.f} $\,\longrightarrow ~$ {\tt fillcoupl2.f}
\item {\tt fillgambr.f} $\,\longrightarrow ~$ {\tt fillgambr2.f}
\end{itemize}
\item New {\tt FORTRAN} files:
\begin{itemize}
\item {\tt filldhpg.f} is to calculate the full propagator matrices 
$D^{H^0\,,H^\pm}(\hat{s})$ and
the $\hat{s}$-dependent couplings $S^{g\,,\gamma}_i(\sqrt{\hat{s}})$
and $P^{g\,,\gamma}_i(\sqrt{\hat{s}})$.
\item {\tt higgsedm.f} is to calculate Higgs-mediated two-loop EDMs of Thallium, electron, and muon.
\item {\tt fillbobs.f} is to calculate the $B$-meson observables:
$B(B_s\to \mu\mu)$, $B(B_d\to \tau\tau)$, 
$\Delta M_{B_d}^{\rm SUSY}$, $\Delta M_{B_s}^{\rm SUSY}$,
$R_{B\tau\nu}$, $B(B\to X_s \gamma)$, and ${\cal A}_{\rm CP}(B\to X_s\gamma)$.
\end{itemize}
\item New flags:
\begin{itemize}
\item ${\tt IFLAG\_H(12)=0-5}$: For the level of improvement in the calculation of the
Higgs-boson pole masses.
\item ${\tt IFLAG\_H(13)=1}$: Not to include the off-diagonal absorptive parts in the
propagator matrices $D^{H^0\,,H^\pm}(\hat{s})$.
\item ${\tt IFLAG\_H(14)=1}$: Print out the the elements of the
full propagator matrices $D^{H^0\,,H^\pm}(\hat{s})$ and
the $\hat{s}$-dependent couplings $S^{g\,,\gamma}_i(\sqrt{\hat{s}})$
and $P^{g\,,\gamma}_i(\sqrt{\hat{s}})$.
\item ${\tt IFLAG\_H(15)=1}$: Print out EDMs.
\item ${\tt IFLAG\_H(16)=1}$: Print out $B$-meson observables.
\item ${\tt IFLAG\_H(17)=1}$: Print out $B \to X_s\,\gamma$ details.
\item ${\tt IFLAG\_H(57)=1}$: This is an error message that appears 
when one of the magnitudes of the complex SUSY input parameters is negative.
\item ${\tt IFLAG\_H(60)=1}$: This is an error message that appears 
when the iterative method for the neutral Higgs-boson pole masses fails.
\end{itemize}
\end{itemize}

\section{Goldstone-boson couplings to third-generation fermions and sfermions}
\label{sec:Goldstone}
Here we present the Goldstone-(s)fermion-(s)fermion couplings in the {\tt CPsuperH}
convention.
\begin{itemize}
\item[$\bullet$] \underline{$G^0$-$\bar{f}$-$f$}
\begin{eqnarray}
{\cal L}_{G^0\bar{f}f}=-\sum_{f=t,b,\tau}\frac{g\,m_f}{2M_W}\,G^0\,\bar{f}
\left(i\, g^P_{G^0\bar{f}f}\, \gamma_5\right)f\,,
\end{eqnarray}
where
\begin{equation}
g^P_{G^0\bar{t}t}=-1\,, \ \ \
g^P_{G^0\bar{b}b}=g^P_{G^0\bar{\tau}\tau}=+1\,.
\end{equation}
\item[$\bullet$] \underline{$G^\pm$-$\bar{f}$-$f^\prime$}
\begin{eqnarray}
{\cal L}_{G^\pm\bar{f}f^\prime}& =& \frac{g}{\sqrt{2} M_W}\, \hskip -0.2cm
\sum_{(f_{\uparrow},f_{\downarrow})=(t,b),(\nu,\tau)} \hskip -0.3cm
G^+\, \bar{f}_{\uparrow}\,\Big(\, m_{f_{\uparrow}}\,
P_L\ -\ m_{f_{\downarrow}}\, P_R\, \Big)\, f_{\downarrow}\ +\ {\rm h.c.} \\
&=&
-g_{tb}\,
G^+\,\bar{t}\,(g^S_{G^+\bar{t}b}+ig^P_{G^+\bar{t}b}\gamma_5)\,b
-g_{\nu_\tau\tau}\,
G^+\,\bar{\nu}_\tau\,(g^S_{G^+\bar{\nu}_\tau\tau}+ig^P_{G^+\bar{\nu}_\tau\tau}\gamma_5)\,\tau
\ +\ {\rm h.c.}\,,\nonumber
\end{eqnarray}
where
\begin{eqnarray}
&&g_{tb}=-\frac{g\,m_t}{\sqrt{2} M_W}\,, \ \ \
g^S_{G^+\bar{t}b}=\frac{1-m_b/m_t}{2}\,, \ \ \
g^P_{G^+\bar{t}b}=i\,\frac{1+m_b/m_t}{2}\,; \nonumber \\
&&g_{\nu_\tau\tau}=-\frac{g\,m_\tau}{\sqrt{2} M_W}\,, \ \ 
g^S_{G^+\bar{\nu}_\tau\tau}=-\frac{1}{2}\,, \hspace{1.50cm}
g^P_{G^+\bar{\nu}_\tau\tau}=i\,\frac{1}{2}\,.
\end{eqnarray}
\item[$\bullet$] \underline{$G^0$-$\tilde{f}^*$-$\tilde{f}$}
\begin{equation}
{\cal L}_{G^0\tilde{f}\tilde{f}}=v\sum_{f=t,b,\tau}\,g_{G^0\tilde{f}^*_i\tilde{f}_j}
(G^0\,\tilde{f}^*_i\,\tilde{f}_j)\,,
\end{equation}
where
\begin{equation}
v\,g_{G^0\tilde{f}^*_i\tilde{f}_j}
=\left(\Gamma^{G^0\tilde{f}^*\tilde{f}}\right)_{\alpha\beta}
U^{\tilde{f}*}_{\alpha i} U^{\tilde{f}}_{\beta j}\,.
\end{equation}
The couplings in the weak-interaction basis are given by
\begin{eqnarray}
\Gamma^{G^0\tilde{t}^*\tilde{t}} &=& \frac{1}{\sqrt{2}}\left(
\begin{array}{cc}
0 & i\,h_t^*(s_\beta A_t^*-c_\beta \mu) \\
-i\,h_t(s_\beta A_t-c_\beta \mu^*) & 0
\end{array} \right)\,,
\nonumber \\
\Gamma^{G^0\tilde{b}^*\tilde{b}} &=& \frac{1}{\sqrt{2}}\left(
\begin{array}{cc}
0 & -i\,h_b^*(c_\beta A_b^*-s_\beta \mu) \\
i\,h_b(c_\beta A_b-s_\beta \mu^*) & 0
\end{array} \right)\,,
\nonumber \\
\Gamma^{G^0\tilde{\tau}^*\tilde{\tau}} &=& \frac{1}{\sqrt{2}}\left(
\begin{array}{cc}
0 & -i\,h_\tau^*(c_\beta A_\tau^*-s_\beta \mu) \\
i\,h_\tau(c_\beta A_\tau-s_\beta \mu^*) & 0
\end{array} \right)\,.
\end{eqnarray}
\item[$\bullet$] \underline{$G^\pm$-$\tilde{f}^*$-$\tilde{f}^\prime$}
\begin{equation}
{\cal L}_{G^\pm\tilde{f}\tilde{f'}}=v\,g_{G^+\tilde{t}^*_i\tilde{b}_j}
(G^+\,\tilde{t}^*_i\,\tilde{b}_j)\,
+\,v\,g_{G^+\tilde{\nu}_\tau^*\tilde{\tau}_i}
(G^+\,\tilde{\nu}_\tau^*\,\tilde{\tau}_i)\,+{\rm h.c.}\,,
\end{equation}
where
\begin{equation}
v\,g_{G^+\tilde{t}^*_i\tilde{b}_j}
=\left(\Gamma^{G^+\tilde{t}^*\tilde{b}}\right)_{\alpha\beta}
U^{\tilde{t}*}_{\alpha i} U^{\tilde{b}}_{\beta j} \ \ \ \  {\rm and} \ \ \ \
v\,g_{G^+\tilde{\nu}^*_\tau\tilde{\tau}_i}
=\Gamma^{G^+\tilde{\nu}^*_\tau\tilde{\tau}_\alpha}\,
U^{\tilde{\tau}}_{\alpha i}\,.
\end{equation}
The couplings in the weak-interaction basis are given by
\begin{eqnarray}
\Gamma^{G^+\tilde{t}^*\tilde{b}}\ &=&\ \left(
\begin{array}{cc} \frac{1}{\sqrt{2}}\,(|h_u|^2 s_\beta^2 - |h_d|^2 c_\beta^2)\,v \,
+\,\frac{1}{2\sqrt{2}}\,g^2c_{2\beta}\,v
&- h_d^*\, ( c_\beta A^*_d - s_\beta \mu )\\
h_u\,( s_\beta A_u - c_\beta \mu^*) &
0 \end{array}\right)\,,
\nonumber\\
\Gamma^{G^+\tilde{\nu}_\tau^*\tilde{\tau}_L}\ &=& \
 -\frac{1}{\sqrt{2}}\,|h_\tau|^2c_\beta^2\,v\,+\,\frac{1}{2\sqrt{2}}\,g^2c_{2\beta}\,v\,,
\nonumber \\
\Gamma^{G^+\tilde{\nu}_\tau^*\tilde{\tau}_R}\ &=& \
  -h^*_\tau \left(c_\beta A^*_\tau-s_\beta \mu\right)\,.
\end{eqnarray}
\end{itemize}
\setcounter{equation}{0}
\section{Sample new outputs}
Here we show the new outputs of {\tt CPsuperH2.0} for the CPX scenario
with $\tan\beta=5$, $M_{H^\pm}=300$ GeV, $M_{\rm SUSY}=500$ GeV, and $\Phi_A=\Phi_3=90^\circ$.
\begin{itemize}
\item ${\tt IFLAG\_H(1)} =1$: In the new version, we are using $m_b(m_t^{\rm pole})=3.155$ GeV
and $m_c(m_t^{\rm pole})=0.735$ GeV as defaults. Note also that the list of the SM
and SUSY input parameters is
extended to include the CKM matrix and the diagonal sfermion mass matrices.
\\
{\tt
~---------------------------------------------------------\\
$~~$Standard~Model~Parameters~~in~/HC\_SMPARA/\\
~---------------------------------------------------------\\
$~~$AEM\_H~~~~=~0.7812E-02~:~alpha\_em(MZ)\\
$~~$ASMZ\_H~~~=~0.1185E+00~:~alpha\_s(MZ)\\
$~~$MZ\_H~~~~~=~0.9119E+02~:~Z~boson~mass~in~GeV\\
$~~$SW\_H~~~~~=~0.4808E+00~:~sinTheta\_W\\
$~~$ME\_H~~~~~=~0.5000E-03~:~electron~mass~in~GeV\\
$~~$MMU\_H~~~~=~0.1065E+00~:~muon~mass~in~GeV\\
$~~$MTAU\_H~~~=~0.1777E+01~:~tau~mass~in~GeV\\
$~~$MDMT\_H~~~=~0.4000E-02~:~d-quark~mass~at~M\_t\^{}pole~in~GeV\\
$~~$MSMT\_H~~~=~0.9000E-01~:~s-quark~mass~at~M\_t\^{}pole~in~GeV\\
$~~$MBMT\_H~~~=~0.3155E+01~:~b-quark~mass~at~M\_t\^{}pole~in~GeV\\
$~~$MUMT\_H~~~=~0.2000E-02~:~u-quark~mass~at~M\_t\^{}pole~in~GeV\\
$~~$MCMT\_H~~~=~0.7350E+00~:~c-quark~mass~at~M\_t\^{}pole~in~GeV\\
$~~$MTPOLE\_H~=~0.1743E+03~:~t-quark~pole~mass~in~GeV\\
$~~$GAMW\_H~~~=~0.2118E+01~:~Gam\_W~in~GeV\\
$~~$GAMZ\_H~~~=~0.2495E+01~:~Gam\_Z~in~GeV\\
$~~$EEM\_H~~~~=~0.3133E+00~:~e~=~(4*pi*alpha\_em)\^{}1/2\\
$~~$ASMT\_H~~~=~0.1084E+00~:~alpha\_s(M\_t\^{}pole)\\
$~~$CW\_H~~~~~=~0.8768E+00~:~cosTheta\_W\\
$~~$TW\_H~~~~~=~0.5483E+00~:~tanTheta\_W\\
$~~$MW\_H~~~~~=~0.7996E+02~:~W~boson~mass~MW~=~MZ*CW\\
$~~$GW\_H~~~~~=~0.6517E+00~:~SU(2)~gauge~coupling~~gw=e/s\_W\\
$~~$GP\_H~~~~~=~0.3573E+00~:~U(1)\_Y~gauge~coupling~gp=e/c\_W\\
$~~$V\_H~~~~~~=~0.2454E+03~:~V~=~2~MW~/~gw\\
$~~$GF\_H~~~~~=~0.1174E-04~:~GF=sqrt(2)*gw\^{}2/8~MW\^{}2~in~GeV\^{}-2\\
$~~$MTMT\_H~~~=~0.1666E+03~:~t-quark~mass~at~M\_t\^{}pole~in~GeV\\
~---------------------------------------------------------\\
$~~$CKM~Matrix~:\\
$~~$|V\_ud|~~~=~|(0.9738E+00~0.0000E+00)|~=~0.9738E+00\\
$~~$|V\_us|~~~=~|(0.2272E+00~0.0000E+00)|~=~0.2272E+00\\
$~~$|V\_ub|~~~=~|(0.2174E-02~-.3349E-02)|~=~0.3993E-02\\
$~~$|V\_cd|~~~=~|(-.2271E+00~-.1377E-03)|~=~0.2271E+00\\
$~~$|V\_cs|~~~=~|(0.9730E+00~-.3213E-04)|~=~0.9730E+00\\
$~~$|V\_cb|~~~=~|(0.4222E-01~0.0000E+00)|~=~0.4222E-01\\
$~~$|V\_td|~~~=~|(0.7478E-02~-.3259E-02)|~=~0.8157E-02\\
$~~$|V\_ts|~~~=~|(-.4161E-01~-.7602E-03)|~=~0.4162E-01\\
$~~$|V\_tb|~~~=~|(0.9991E+00~0.0000E+00)|~=~0.9991E+00\\
~---------------------------------------------------------\\
$~~$Real~SUSY~Parameters~~in~/HC\_RSUSYPARA/\\
~---------------------------------------------------------\\
$~~$TB\_H~~~~~=~0.5000E+01~:~tan(beta)\\
$~~$CB\_H~~~~~=~0.1961E+00~:~cos(beta)\\
$~~$SB\_H~~~~~=~0.9806E+00~:~sin(beta)\\
$~~$MQ3\_H~~~~=~0.5000E+03~:~M\_tilde{Q\_3}~in~GeV\\
$~~$MU3\_H~~~~=~0.5000E+03~:~M\_tilde{U\_3}~in~GeV\\
$~~$MD3\_H~~~~=~0.5000E+03~:~M\_tilde{D\_3}~in~GeV\\
$~~$ML3\_H~~~~=~0.5000E+03~:~M\_tilde{L\_3}~in~GeV\\
$~~$ME3\_H~~~~=~0.5000E+03~:~M\_tilde{E\_3}~in~GeV\\
~---------------------------------------------------------\\
$~~$Complex~SUSY~Parameters~~in~/HC\_CSUSYPARA/\\
~---------------------------------------------------------\\
$~~$|MU\_H|~~~~~=~0.2000E+04:Mag.~of~MU~parameter~in~GeV\\
$~~$|M1\_H|~~~~~=~0.5000E+02:Mag.~of~M1~parameter~in~GeV\\
$~~$|M2\_H|~~~~~=~0.1000E+03:Mag.~of~M2~parameter~in~GeV\\
$~~$|M3\_H|~~~~~=~0.1000E+04:Mag.~of~M3~parameter~in~GeV\\
$~~$|AT\_H|~~~~~=~0.1000E+04:Mag.~of~AT~parameter~in~GeV\\
$~~$|AB\_H|~~~~~=~0.1000E+04:Mag.~of~AB~parameter~in~GeV\\
$~~$|ATAU\_H|~~~=~0.1000E+04:Mag.~of~ATAU~parameter~in~GeV\\
$~~$ARG(MU\_H)~~=~0.0000E+00:Arg.~of~MU~parameter~in~Degree\\
$~~$ARG(M1\_H)~~=~0.0000E+00:Arg.~of~M1~parameter~in~Degree\\
$~~$ARG(M2\_H)~~=~0.0000E+00:Arg.~of~M2~parameter~in~Degree\\
$~~$ARG(M3\_H)~~=~0.9000E+02:Arg.~of~M3~parameter~in~Degree\\
$~~$ARG(AT\_H)~~=~0.9000E+02:Arg.~of~AT~parameter~in~Degree\\
$~~$ARG(AB\_H)~~=~0.9000E+02:Arg.~of~AB~parameter~in~Degree\\
$~~$ARG(ATAU\_H)=~0.9000E+02:Arg.~of~ATAU~parameter~in~Degree\\
~---------------------------------------------------------\\
$~~$Diagonal~Sfermion~Mass~Matrices~[GeV]~(Not~squared)~:\\
$~~$M\_Q~=~0.5000E+03~x~Diag(0.1000E+01~0.1000E+01~0.1000E+01)\\
$~~$M\_U~=~0.5000E+03~x~Diag(0.1000E+01~0.1000E+01~0.1000E+01)\\
$~~$M\_D~=~0.5000E+03~x~Diag(0.1000E+01~0.1000E+01~0.1000E+01)\\
$~~$M\_L~=~0.5000E+03~x~Diag(0.1000E+01~0.1000E+01~0.1000E+01)\\
$~~$M\_E~=~0.5000E+03~x~Diag(0.1000E+01~0.1000E+01~0.1000E+01)\\
~---------------------------------------------------------\\
$~~$Charged~Higgs~boson~pole~mass~:~0.3000E+03~GeV\\
~---------------------------------------------------------\\
} 
\item ${\tt IFLAG\_H(2)}=1$: The masses and mixing matrix of the neutral Higgs boson change 
due to the improvement in their calculations and the new input for the $b$-quark mass.\\
{\tt
~---------------------------------------------------------\\
$~~$Masses~and~Mixing~Matrix~of~Higgs~bosons~:\\
$~~~~~~~~~~~~~~~~~~~~~~~~~~~~~~~~$HMASS\_H(I)~and~OMIX\_H(A,I)\\
~---------------------------------------------------------\\
$~~$H1~~Pole~Mass~~~~~~~~~~~=~0.1193E+03~GeV\\
$~~$H2~~Pole~Mass~~~~~~~~~~~=~0.2718E+03~GeV\\
$~~$H3~~Pole~Mass~~~~~~~~~~~=~0.2983E+03~GeV\\
$~~$Charged~Higgs~Pole~Mass~=~0.3000E+03~GeV~[SSPARA\_H(2)]\\
$~~~~~~~~~~~~~~~~~~~~~~~~~~$[H1]~~~~~~~~[H2]~~~~~~~~[H3]\\
$~~~~~~~~~~~~$[phi\_1]~/~0.2457E+00~~0.3360E+00~~0.9093E+00~~$\backslash$\\
$~~$O(IA,IH)=~[phi\_2]~|~0.9693E+00~~-.7551E-01~~-.2340E+00~~|\\
$~~~~~~~~~~~~$[~~a~~]~$\backslash$~-.9973E-02~~0.9388E+00~~-.3442E+00~~/\\
 ---------------------------------------------------------\\
} 
\item ${\tt IFLAG\_H(14)}=1$: The elements of the propagator matrices  
$D^{H^0\,,H^\pm}(\hat{s})$
and the $\hat{s}$-dependent couplings of the neutral Higgs bosons 
to two photons, $S^{\gamma}_i(\sqrt{\hat{s}})$ and $P^{\gamma}_i(\sqrt{\hat{s}})$, 
and two gluons, $S^{g}_i(\sqrt{\hat{s}})$ and $P^{g}_i(\sqrt{\hat{s}})$, taking 
$\sqrt{\hat{s}}=M_{H_2}$. The couplings are compared to their values at the Higgs-boson
pole masses:
$S^{\gamma}_i(\sqrt{\hat{s}}=M_{\tt IH})={\tt NHC\_H(88,IH)}$,
$P^{\gamma}_i(\sqrt{\hat{s}}=M_{\tt IH})={\tt NHC\_H(89,IH)}$,
$S^g_i(\sqrt{\hat{s}}=M_{\tt IH})={\tt NHC\_H(84,IH)}$,
$P^g_i(\sqrt{\hat{s}}=M_{\tt IH})={\tt NHC\_H(85,IH)}$.
\\
{\tt
~---------------------------------------------------------\\
$~~$DNH4~at~sqrt{s}~=~0.2718E+03~GeV\\
~---------------------------------------------------------\\
$~~$DNH4[H1,H1]:~|(0.1238E+01~0.2290E-01)|~=~0.1238E+01\\
$~~$DNH4[H2,H2]:~|(0.5542E-01~-.1611E+04)|~=~0.1611E+04\\
$~~$DNH4[H3,H3]:~|(-.4876E+01~-.2128E-01)|~=~0.4876E+01\\
$~~$DNH4[H1,H2]:~|(-.1607E+00~-.2973E-02)|~=~0.1608E+00\\
$~~$DNH4[H1,H3]:~|(-.5956E-05~0.2606E-03)|~=~0.2606E-03\\
$~~$DNH4[H2,H3]:~|(-.4893E-01~-.2377E-03)|~=~0.4893E-01\\
$~~$DNH4[G0,H1]:~|(0.3403E-06~-.1825E-04)|~=~0.1825E-04\\
$~~$DNH4[G0,H2]:~|(0.1872E+00~0.2446E-05)|~=~0.1872E+00\\
$~~$DNH4[G0,H3]:~|(-.2222E-06~0.5181E-04)|~=~0.5182E-04\\
$~~$DNH4[G0,G0]:~|(0.1000E+01~-.2365E-05)|~=~0.1000E+01\\
~---------------------------------------------------------\\
$~~$DCH2~at~sqrt{s}~=~0.2718E+03~GeV\\
~---------------------------------------------------------\\
$~~$DCH2[H+,H+]:~|(-.4576E+01~-.2790E-01)|~=~0.4576E+01\\
$~~$DCH2[H+,G+]:~|(-.1256E-03~0.2294E-01)|~=~0.2294E-01\\
$~~$DCH2[G+,H+]:~|(-.1256E-03~0.2294E-01)|~=~0.2294E-01\\
$~~$DCH2[G+,G+]:~|(0.1000E+01~-.6202E-03)|~=~0.1000E+01\\
~---------------------------------------------------------\\
$~~$Comparisons~of~the~H-photon-photon~couplings~at~MH\^{}pole\\
$~~$and~those~at~sqrt\{s\}~=~0.2718E+03~GeV\\
~---------------------------------------------------------\\
$~~~~~~~~~~~~~~~~~~$S~couplings~~~~~~~~~~~~~P~couplings\\
$~~$H1PP(M):~(-.6615E+01~0.6386E-01)~(0.1303E-01~0.7314E-03)\\
$~~$H1PP(S):~(-.3180E+01~-.6078E+01)~(0.1779E-01~0.2017E-02)\\
$~~$H2PP(M):~(-.9852E+00~0.3333E-01)~(-.6867E+00~-.2221E+00)\\
$~~$H2PP(S):~(-.9852E+00~0.3333E-01)~(-.6867E+00~-.2221E+00)\\
$~~$H3PP(M):~(-.4272E+00~0.2509E+00)~(0.5178E+00~0.7028E-01)\\
$~~$H3PP(S):~(-.3695E+00~0.2852E+00)~(0.4567E+00~0.7475E-01)\\
~---------------------------------------------------------\\
$~~$Comparisons~of~the~H-glue-glue~couplings~at~MH\^{}pole\\
$~~$and~those~at~sqrt\{s\}~=~0.2718E+03~GeV\\
~---------------------------------------------------------\\
$~~~~~~~~~~~~~~~~~~$S~couplings~~~~~~~~~~~~~P~couplings\\
$~~$H1GG(M):~(0.5792E+00~0.4164E-01)~(0.5316E-02~-.6809E-03)\\
$~~$H1GG(S):~(0.7358E+00~0.8932E-02)~(0.6510E-02~-.1457E-03)\\
$~~$H2GG(M):~(-.3557E+00~0.2591E-02)~(-.1970E+00~-.3456E-01)\\
$~~$H2GG(S):~(-.3557E+00~0.2591E-02)~(-.1970E+00~-.3456E-01)\\
$~~$H3GG(M):~(-.2240E+00~0.2860E-01)~(0.1855E+00~0.2231E-02)\\
$~~$H3GG(S):~(-.2150E+00~0.3413E-01)~(0.1585E+00~0.2662E-02)\\
~---------------------------------------------------------\\
} 
\item ${\tt IFLAG\_H(15)}=1$: The Higgs-mediated two-loop Thallium, electron, and muon EDMs. For
the Thallium case, the two main contributions from the electron EDM and the CP-odd
electron-nucleon interaction are shown separately.\\
{\tt
~---------------------------------------------------------\\
$~~~~~~~~~~~~~~~$Higgs-mediated~two-loop~EDMs\\
$~~~~~~~$Phi\_3~=~0.9000E+02\^{}o~and~Phi\_At~=~0.9000E+02\^{}o\\
~---------------------------------------------------------\\
$~~$Thallium[10\^{}-24~ecm]:~-.2612E+01\\
$~~~~~~~~~~~~~~~~~~~~~~~$[-.2568E+01~from~electron~EDM]\\
$~~~~~~~~~~~~~~~~~~~~~~~$[-.4467E-01~from~C\_S\,~~~~~~EDM]\\
$~~$Electron[10\^{}-26~ecm]:~0.4389E+00\\
$~~$Muon[10\^{}-24~ecm]~~~~:~0.8997E+00\\
~---------------------------------------------------------\\
} 
\item ${\tt IFLAG\_H(16)}=1$: The $B$-meson observables.\\
{\tt
~---------------------------------------------------------\\
$~~~~~~~~~~~~~~~~~~~~~~$B~Observables\\
~---------------------------------------------------------\\
$~~$B(B\_s~->~mu~~mu~)~~~x~10\^{}7~=~0.3710E-01\\
$~~$B(B~~~->~X\_s~gamma)~x~10\^{}4~=~0.4396E+01\\
$~~$B(B\_u~->~tau~nu)/B(SM)~~~~~=~0.9854E+00\\
$~~$B(B\_d~->~tau~tau)~~~x~10\^{}7~=~0.2294E+00\\
$~~$ACP(B~->~X\_s~gamma)~x~10\^{}2~=~-.7954E-01~[\%]\\
$~~$Delta~M~[B\_d]~(SUSY)~~~~~~~=~0.6659E-04~[1/ps]\\
$~~$Delta~M~[B\_s]~(SUSY)~~~~~~~=~0.1982E-01~[1/ps]\\
~---------------------------------------------------------\\
} 
\item ${\tt IFLAG\_H(17)}=1$: The details of the $B \to X_s \gamma$ calculation. As a default,
we use $m_c(\mu_c=m_c^{\rm pole})$ to
capture a part of NNLO corrections \cite{b2sg-nnlo}. The case when only the
charged-Higgs contribution is added to the SM prediction is also shown.
\\
{\tt
~---------------------------------------------------------\\
$~~~~~~~~~~~~~~~~~~~~~~$B~->~X\_s~gamma\\
$~~$delta~and~E\_gamma\^{}cut~[GeV]:~0.3333E+00~~0.1601E+01\\
~---------------------------------------------------------\\
$~~$b-q~masses~[GeV]~~(pole,~~~~~~~@mb\^{}pole,~~~@mt\^{}pole):\\
$~~~~~~~~~~~~~~~~~~~~~$0.4802E+01~~0.4415E+01~~0.3155E+01\\
$~~$c-q~masses~[GeV]~~(pole,~~~~~~~@mc\^{}pole,~~~@mb\^{}pole):\\
$~~~~~~~~~~~~~~~~~~~~~$0.1415E+01~~0.1250E+01~~0.1029E+01\\
$~~$mu\_b~and~mu\_c~~[GeV]~~~~~~~:~0.4802E+01~~0.1415+01\\
~---------------------------------------------------------\\
$~~$BR~~x~10\^{}4:~0.4396E+01~(SM+Charged~Higgs+Chargino)\\
$~~~~~~~~~~~~~$[0.4471E+01~(SM+Charged~Higgs)]\\
$~~~~~~~~~~~~~$[0.3351E+01~(SM)]\\
$~~$ACP~x~10\^{}2:~-.7954E-01~\%\\
~---------------------------------------------------------\\
} 
\end{itemize}

\end{appendix}

\newpage
%
%

\begin{table}[\hbt]
\caption{\label{tab:raux}
{\it
The contents of the array {\tt RAUX\_H}. In {\tt RAUX\_H(22)} and {\tt RAUX\_H(23)},
the notation $h_f^0$ is for the Yukawa couplings without including the threshold
corrections. The notations which are not explained in the text 
follow the conventions of {\tt CPsuperH} \cite{cpsuperh}  and
Refs.~\cite{Carena:2000yi,Carena:2001fw,MCPMFV}.
}}
\begin{center}
\begin{tabular}{|cl|cl|cl|}
\hline
{\tt RAUX\_H(1)} & $m_b^{\rm pole}$ & 
{\tt RAUX\_H(26)} & $\!\!\! |h_t(Q_{tb})|$ & 
{\tt RAUX\_H(120)} &  $d^H_\mu\times 10^{24}\,e\,cm$ \\ 
{\tt RAUX\_H(2)} & $m_b(m_b^{\rm pole})$ & 
{\tt RAUX\_H(27)} & $\!\!\! |h_b(m_t^{\rm pole})|$ & 
... &  ... \\ 
{\tt RAUX\_H(3)} & $\alpha_s(m_b^{\rm pole})$ & 
{\tt RAUX\_H(28)} & $\!\!\! |h_b(Q_b)|$ & 
... &  ... \\ 
{\tt RAUX\_H(4)} & $m_c^{\rm pole}$ & 
{\tt RAUX\_H(29)} & $\!\!\! |h_b(Q_{tb})|$ & 
... & ... \\
{\tt RAUX\_H(5)} & $m_c(m_c^{\rm pole})$ & 
{\tt RAUX\_H(30)} & $\!\!\! M_A^2$ & 
... & ... \\
{\tt RAUX\_H(6)} & $\alpha_s(m_c^{\rm pole})$ & 
{\tt RAUX\_H(31)} & $\!\!\!\!\real\widehat{\Pi}_{H^+H^-}(M_{H^\pm}^{{\rm pole}\,2})$ & 
... & ... \\
... & ... & 
{\tt RAUX\_H(32)} & $\!\!\! \bar\lambda_4v^2(m_t^{\rm pole})/2$ & 
... & ... \\
... & ... & 
{\tt RAUX\_H(33)} &  $\!\!\! \bar\lambda_4(m_t^{\rm pole})$ & 
{\tt RAUX\_H(130)} & $\!\!\!\!B(B_s\!\to\!\mu \mu)\!\times\!\! 10^7$ \\
... & ...  & 
{\tt RAUX\_H(34)} & $\!\!\! \bar\lambda_1(m_t^{\rm pole})$ & 
{\tt RAUX\_H(131)} & $\!\!\!\!B(B_d\!\to\!\tau \tau)\!\times\!\! 10^7$ \\
{\tt RAUX\_H(10)} & $M_{H^\pm}^{\rm pole}$ or $M_{H^\pm}^{\rm eff.}$ & 
{\tt RAUX\_H(35)} & $\!\!\! \bar\lambda_2(m_t^{\rm pole})$ & 
{\tt RAUX\_H(132)} & $\!\!\!\Delta M_{B_d}^{\rm SUSY}\,{\rm ps}^{-1}$ \\
{\tt RAUX\_H(11)} & $Q_t^2$  & 
{\tt RAUX\_H(36)} & $\!\!\! \bar\lambda_{34}(m_t^{\rm pole})$ &
{\tt RAUX\_H(133)} & $\!\!\!\Delta M_{B_s}^{\rm SUSY}\,{\rm ps}^{-1}$ \\
{\tt RAUX\_H(12)} & $Q_b^2$  & 
... & ... & 
{\tt RAUX\_H(134)} & $\!\!\!R_{B\tau\nu}$ \\
{\tt RAUX\_H(13)} & $Q_{tb}^2$ & 
... & ... & 
{\tt RAUX\_H(135)} & $\!\!\!\!B(B\!\to\!X_s \gamma)\!\times\!\! 10^4$ \\
{\tt RAUX\_H(14)} & $v_1(m_t^{\rm pole})$ & 
{\tt RAUX\_H(101)} & $\!\!\! \sqrt{\hat{s}}$ & 
{\tt RAUX\_H(136)} & $\!\!\!\!{\cal A}_{\rm CP}(B\!\to\!X_s \gamma)\,\%$ \\
{\tt RAUX\_H(15)} & $v_1(Q_t)$ & 
... & ... & 
... & ... \\
{\tt RAUX\_H(16)} & $v_1(Q_b)$ & 
... & ... & 
... & ... \\
{\tt RAUX\_H(17)} & $v_1(Q_{tb})$ & 
{\tt RAUX\_H(111)} & $\!\!\!d^H_{\rm Tl}\times 10^{24}\,e\,cm$ & 
... & ... \\
{\tt RAUX\_H(18)} & $v_2(m_t^{\rm pole})$ & 
{\tt RAUX\_H(112)} &  $\!\!\!(d^H_{\rm Tl})^{\,e}\times 10^{24}\,e\,cm$ & 
& \\
{\tt RAUX\_H(19)} & $v_2(Q_t)$ & 
{\tt RAUX\_H(113)} &  $\!\!\!(d^H_{\rm Tl})^{C_S}\!\times\! 10^{24}\,e\,cm$ & 
& \\
{\tt RAUX\_H(20)} & $v_2(Q_b)$ & 
{\tt RAUX\_H(114)} &  $\!\!\!d^H_e\times 10^{26}\,e\,cm$ & 
& \\
{\tt RAUX\_H(21)} & $v_2(Q_{tb})$ & 
{\tt RAUX\_H(115)} & $\!\!\!(d^H_e)^{\tilde{t}}\times 10^{26}\,e\,cm$ & 
& \\
{\tt RAUX\_H(22)} & $|h_t^0(m_t^{\rm pole})|$ & 
{\tt RAUX\_H(116)} & $\!\!\!(d^H_e)^{\tilde{b}}\times 10^{26}\,e\,cm$ & 
& \\
{\tt RAUX\_H(23)} & $|h_b^0(m_t^{\rm pole})|$ & 
{\tt RAUX\_H(117)} & $\!\!\!(d^H_e)^t\times 10^{26}\,e\,cm$ & 
& \\
{\tt RAUX\_H(24)} & $|h_t(m_t^{\rm pole})|$ & 
{\tt RAUX\_H(118)} & $\!\!\!(d^H_e)^{b}\times 10^{26}\,e\,cm$ & 
& \\
{\tt RAUX\_H(25)} & $|h_t(Q_t)|$ & 
{\tt RAUX\_H(119)} & $\!\!\!(d^H_e)^{\tilde{\chi}^\pm}\!\times\! 10^{26}\,e\,cm$ & 
& \\
\hline
\end{tabular}
\end{center}
\end{table}
\begin{table}[\hbt]
\caption{\label{tab:caux}
{\it
The contents of the array {\tt CAUX\_H}. 
The notations which are not explained in the text 
follow the {\tt CPsuperH} \cite{cpsuperh} convention.
}}
\begin{center}
\begin{tabular}{|cl|cl|cl|}
\hline
{\tt CAUX\_H(1)} & $h_t/|h_t|$ & 
{\tt CAUX\_H(112)} & $D^{H^0}_{4,1}({\hat{s}})$ & 
{\tt CAUX\_H(140)} & $S^g_1(\sqrt{\hat{s}})$ \\
{\tt CAUX\_H(2)} & $h_b/|h_b|$ & 
{\tt CAUX\_H(113)} & $D^{H^0}_{4,2}({\hat{s}})$ & 
{\tt CAUX\_H(141)} & $P^g_1(\sqrt{\hat{s}})$ \\
... & ... & 
{\tt CAUX\_H(114)} & $D^{H^0}_{4,3}({\hat{s}})$ & 
{\tt CAUX\_H(142)} & $S^g_2(\sqrt{\hat{s}})$ \\
... & ... & 
{\tt CAUX\_H(115)} & $D^{H^0}_{4,4}({\hat{s}})$ & 
{\tt CAUX\_H(143)} & $P^g_2(\sqrt{\hat{s}})$ \\
... & ... & 
{\tt CAUX\_H(116)} & $D^{H^\pm}_{H^\pm,H^\pm}({\hat{s}})$ & 
{\tt CAUX\_H(144)} & $S^g_3(\sqrt{\hat{s}})$ \\
... & ... & 
{\tt CAUX\_H(117)} & $D^{H^\pm}_{H^\pm,G^\pm}({\hat{s}})$ & 
{\tt CAUX\_H(145)} & $P^g_3(\sqrt{\hat{s}})$ \\
{\tt CAUX\_H(100)} & $D^{H^0}_{1,1}({\hat{s}})$ & 
{\tt CAUX\_H(118)} & $D^{H^\pm}_{G^\pm,H^\pm}({\hat{s}})$ & 
... & ... \\
{\tt CAUX\_H(101)} & $D^{H^0}_{1,2}({\hat{s}})$ & 
{\tt CAUX\_H(119)} & $D^{H^\pm}_{G^\pm,G^\pm}({\hat{s}})$ & 
... & ...  \\
{\tt CAUX\_H(102)} & $D^{H^0}_{1,3}({\hat{s}})$ & 
... & ... & 
{\tt CAUX\_H(150)} & $\bra{\bar{B}^0_d}\, H_{\rm eff}^{\Delta B=2}\, \ket{B^0_d}_{\rm SUSY}$ \\
{\tt CAUX\_H(103)} & $D^{H^0}_{1,4}({\hat{s}})$ & 
... & ... & 
{\tt CAUX\_H(151)} & $\bra{\bar{B}^0_s}\, H_{\rm eff}^{\Delta B=2}\, \ket{B^0_s}_{\rm SUSY}$ \\
{\tt CAUX\_H(104)} & $D^{H^0}_{2,1}({\hat{s}})$ & 
... &  ... & 
... & ...  \\
{\tt CAUX\_H(105)} & $D^{H^0}_{2,2}({\hat{s}})$ & 
{\tt CAUX\_H(130)} &  $S^\gamma_1(\sqrt{\hat{s}})$ & 
... & ...  \\
{\tt CAUX\_H(106)} & $D^{H^0}_{2,3}({\hat{s}})$ & 
{\tt CAUX\_H(131)} &  $P^\gamma_1(\sqrt{\hat{s}})$ & 
... & ...  \\
{\tt CAUX\_H(107)} & $D^{H^0}_{2,4}({\hat{s}})$ & 
{\tt CAUX\_H(132)} &  $S^\gamma_2(\sqrt{\hat{s}})$ & 
... & ...  \\
{\tt CAUX\_H(108)} & $D^{H^0}_{3,1}({\hat{s}})$ & 
{\tt CAUX\_H(133)} &  $P^\gamma_2(\sqrt{\hat{s}})$ & 
... & ...  \\
{\tt CAUX\_H(109)} & $D^{H^0}_{3,2}({\hat{s}})$ & 
{\tt CAUX\_H(134)} &  $S^\gamma_3(\sqrt{\hat{s}})$ & 
... & ...  \\
{\tt CAUX\_H(110)} & $D^{H^0}_{3,3}({\hat{s}})$ & 
{\tt CAUX\_H(135)} &  $P^\gamma_3(\sqrt{\hat{s}})$ & 
... & ...  \\
{\tt CAUX\_H(111)} & $D^{H^0}_{3,4}({\hat{s}})$ & 
... &  ... & 
... & ...  \\
\hline
\end{tabular}
\end{center}
\end{table}

\begin{table}[\hbt]
\caption{\label{tab:smpara}
{\it
The contents of the extended {\tt SMPARA\_H(IP)}.  }
}
\begin{center}
\begin{tabular}{|c|c|c|c|c|c|c|c|}
\hline
{\tt IP} & Parameter & {\tt IP} & Parameter
& {\tt IP} & Parameter & {\tt IP} & Parameter \\
\hline
   1 & $\alpha^{-1}_{\rm em}(M_Z)$ &
   6 & $m_\mu$                     &
  11 & $m_u (m_t^{\rm pole})$      &
  16 & $\lambda$ \\
   2 & $\alpha_s(M_Z)$             &
   7 & $m_\tau $                   &
  12 & $m_c (m_t^{\rm pole})$      &
  17 & $A$ \\
   3 & $M_Z$                       &
   8 & $m_d (m_t^{\rm pole})$      &
  13 & $m_t^{\rm pole}$            &
  18 & $\bar\rho$ \\
   4 & $\sin^2\theta_W$            &
   9 & $m_s (m_t^{\rm pole})$      &
  14 & $\Gamma_W$                  &
  19 & $\bar\eta$ \\
   5 & $m_e$                       &
  10 & $m_b (m_t^{\rm pole})$      &
  15 & $\Gamma_Z$                  &
  20 & ... \\
\hline
\end{tabular}
\end{center}
\end{table}
\begin{table}[\hbt]
\caption{\label{tab:sspara}
{\it
The contents of the extended {\tt SSPARA\_H(IP)}.  }
}
\begin{center}
\begin{tabular}{|c|c|c|c|c|c|c|c|c|}
\hline
  {\tt IP} & Parameter
& {\tt IP} & Parameter
& {\tt IP} & Parameter
& {\tt IP} & Parameter \\
\hline
   1 & $\tan\beta$                  &
   8 & $\Phi_2$                     &
  15 & $m_{\tilde{E}_3}$            &
  22 & $\rho_{\tilde{Q}}$                      \\
   2 & $M_{H^\pm}^{\rm pole}$       &
   9 & $|M_3|$                      &
  16 & $|A_t|$                      &
  23 & $\rho_{\tilde{U}}$                 \\
   3 & $|\mu|$                      &
  10 & $\Phi_3$                     &
  17 & $\Phi_{A_t}$            &
  24 & $\rho_{\tilde{D}}$                      \\
   4 & $\Phi_\mu$                   &
  11 & $m_{\tilde{Q}_3}$            &
  18 & $|A_b|$            &
  25 & $\rho_{\tilde{L}}$                 \\
   5 & $|M_1|$                      &
  12 & $m_{\tilde{U}_3}$            &
  19 & $\Phi_{A_b}$            &
  26 & $\rho_{\tilde{E}}$                     \\
   6 & $\Phi_1$                     &
  13 & $m_{\tilde{D}_3}$            &
  20 & $|A_\tau|$            &
  27 & ...                     \\
   7 & $|M_2|$                      &
  14 & $m_{\tilde{L}_3}$            &
  21 & $\Phi_{A_\tau}$            &
  28 & ...                     \\
\hline
\end{tabular}
\end{center}
\end{table}


\newpage

\begin{thebibliography}{99}

\bibitem{Nath} For  a recent review, see  T.~Ibrahim and P.~Nath,
 arXiv:0705.2008 [hep-ph].  

\bibitem{BARYO}
  M.~Trodden,
{\it In the Proceedings of 32nd SLAC Summer Institute on Particle Physics (SSI
2004): Natures Greatest Puzzles, Menlo Park, California, 2-13 Aug
2004, pp L018}
  [arXiv:hep-ph/0411301];
%
  M.~Quiros,
  J.\ Phys.\ A  {\bf 40} (2007) 6573;
%
  W.~Buchmuller,
  arXiv:0710.5857 [hep-ph].

\bibitem{cpsuperh}
  J.~S.~Lee, A.~Pilaftsis, M.~Carena, S.~Y.~Choi, M.~Drees, J.~R.~Ellis and C.~E.~M.~Wagner,
  Comput.\ Phys.\ Commun.\  {\bf 156} (2004) 283
  [arXiv:hep-ph/0307377].

\bibitem{feynhiggs}
S.~Heinemeyer, W.~Hollik and G.~Weiglein, {\em
Comp. Phys. Comm.} {\bf 124} 2000 76,
hep-ph/9812320;
  T.~Hahn, S.~Heinemeyer, W.~Hollik, H.~Rzehak and G.~Weiglein,
  arXiv:0710.4891 [hep-ph].

\bibitem{MCPMFV}
  J.~Ellis, J.~S.~Lee and A.~Pilaftsis,
  arXiv:0708.2079 [hep-ph].


\bibitem{Carena:2001fw}
  M.~Carena, J.~R.~Ellis, A.~Pilaftsis and C.~E.~M.~Wagner,
  Nucl.\ Phys.\ B {\bf 625} (2002) 345
  [arXiv:hep-ph/0111245].

\bibitem{Carena:2000yi}
  M.~Carena, J.~R.~Ellis, A.~Pilaftsis and C.~E.~M.~Wagner,
  Nucl.\ Phys.\ B {\bf 586} (2000) 92
  [arXiv:hep-ph/0003180].


\bibitem{Carena:2000ks}
  M.~Carena, J.~R.~Ellis, A.~Pilaftsis and C.~E.~M.~Wagner,
  Phys.\ Lett.\ B {\bf 495} (2000) 155
  [arXiv:hep-ph/0009212].

\bibitem{Ellis:2004fs}
  J.~R.~Ellis, J.~S.~Lee and A.~Pilaftsis,
  Phys.\ Rev.\ D {\bf 70} (2004) 075010
  [arXiv:hep-ph/0404167].

\bibitem{APNPB} A.~Pilaftsis, Nucl.\ Phys.\ B {\bf 504} (1997) 61.

\bibitem{PT} 
  J.~M.~Cornwall and J.~Papavassiliou,
  Phys.\ Rev.\  D {\bf 40} (1989) 3474;\\
J.~Papavassiliou,
  Phys.\ Rev.\  D {\bf 41} (1990) 3179;\\
 D.~Binosi and J.~Papavassiliou,
  Phys.\ Rev.\  D {\bf 66} (2002) 111901;
  J.\ Phys.\ G {\bf 30} (2004)~203.

\bibitem{gluon_fusion}
  H.~M.~Georgi, S.~L.~Glashow, M.~E.~Machacek and D.~V.~Nanopoulos,
  Phys.\ Rev.\ Lett.\  {\bf 40} (1978) 692;
  T.~Inami, T.~Kubota and Y.~Okada,
  Z.\ Phys.\  C {\bf 18} (1983) 69;
  A.~Djouadi, M.~Spira and P.~M.~Zerwas,
  Phys.\ Lett.\  B {\bf 264} (1991) 440;
  M.~Spira, A.~Djouadi, D.~Graudenz and P.~M.~Zerwas,
  Nucl.\ Phys.\  B {\bf 453} (1995) 17
  [arXiv:hep-ph/9504378].

\bibitem{lhc_cp}
  A. Dedes and S. Moretti, Phys.\ Rev.\ Lett.\ {\bf 84}
  (2000) 22; Nucl.\ Phys.\ B {\bf 576} (2000) 29; S.Y. Choi and J.S. Lee,
  Phys.\ Rev.\ D {\bf 61} (2000) 115002; S.Y.  Choi, K.  Hagiwara and
  J.S. Lee, Phys.\ Lett.\ B {\bf 529} (2002) 212;
  A.~Arhrib,  D.~K.~Ghosh and O.C.~Kong,
  Phys.\ Lett.\ B {\bf 537}   (2002)  217;
  E.~Christova, H.~Eberl, W.~Majerotto and S.~Kraml,
  Nucl.\ Phys.\ B {\bf 639} (2002) 263; JHEP {\bf 0212} (2002) 021;
  W.~Khater and P.~Osland, Nucl.\ Phys.\ B {\bf 661} (2003) 209.

\bibitem{photon_collider}
  J.~F.~Gunion and H.~E.~Haber,
  Phys.\ Rev.\  D {\bf 48} (1993) 5109;
  D.~L.~Borden, D.~A.~Bauer and D.~O.~Caldwell,
  Phys.\ Rev.\  D {\bf 48} (1993) 4018;
  B.~Grzadkowski and J.~F.~Gunion,
  Phys.\ Lett.\  B {\bf 294} (1992) 361
  [arXiv:hep-ph/9206262];
  M.~Kramer, J.~H.~Kuhn, M.~L.~Stong and P.~M.~Zerwas,
  Z.\ Phys.\  C {\bf 64} (1994) 21
  [arXiv:hep-ph/9404280];
  G.~J.~Gounaris and G.~P.~Tsirigoti,
  Phys.\ Rev.\  D {\bf 56} (1997) 3030
  [Erratum-ibid.\  D {\bf 58} (1998) 059901]
  [arXiv:hep-ph/9703446];
  I.~F.~Ginzburg, G.~L.~Kotkin, S.~L.~Panfil, V.~G.~Serbo and V.~I.~Telnov,
  Nucl.\ Instrum.\ Meth.\  A {\bf 219} (1984) 5;
  B.~Badelek {\it et al.}  [ECFA/DESY Photon Collider Working Group],
  Int.\ J.\ Mod.\ Phys.\  A {\bf 19} (2004) 5097
  [arXiv:hep-ex/0108012].

\bibitem{photon_cp}
  S.~Y.~Choi and J.~S.~Lee, Phys.\ Rev.\ D {\bf 62}
  (2000) 036005; E.~Asakawa, S.~Y.~Choi, K.~Hagiwara and J.S. Lee,
  Phys.\ Rev.\ D {\bf 62} (2000) 115005; J.~S.~Lee, hep-ph/0106327;
  S.~Y.~Choi, B.~C.~Chung, P.~Ko and J.~S.~Lee, Phys.\ Rev.\ D {\bf 66}
  (2002) 016009; R.~M.~Godbole, S.~D.~Rindani and R.~K.~Singh,
  Phys.\ Rev.\ D {\bf 67} (2003) 095009;
  E.~Asakawa and K.~Hagiwara,
  Eur.\ Phys.\ J.\  C {\bf 31} (2003) 351
  [arXiv:hep-ph/0305323];
  S.~Y.~Choi, J.~Kalinowski, Y.~Liao and P.~M.~Zerwas,
  Eur.\ Phys.\ J.\  C {\bf 40} (2005) 555
  [arXiv:hep-ph/0407347];
  B.~Grzadkowski, Z.~Hioki, K.~Ohkuma and J.~Wudka,
  JHEP {\bf 0511} (2005) 029
  [arXiv:hep-ph/0508183];
  R.~M.~Godbole, S.~Kraml, S.~D.~Rindani and R.~K.~Singh,
  Phys.\ Rev.\  D {\bf 74} (2006) 095006
  [Erratum-ibid.\  D {\bf 74} (2006) 119901]
  [arXiv:hep-ph/0609113].

\bibitem{Ellis:2004hw}
  J.~R.~Ellis, J.~S.~Lee and A.~Pilaftsis,
  Nucl.\ Phys.\  B {\bf 718} (2005) 247
  [arXiv:hep-ph/0411379];
  J.~S.~Lee,
  Mod.\ Phys.\ Lett.\  A {\bf 22} (2007) 1191
  [arXiv:0705.1089 [hep-ph]].

\bibitem{QCD1}
  M.~Spira, A.~Djouadi, D.~Graudenz and P.~M.~Zerwas,
  Phys.\ Lett.\  B {\bf 318} (1993) 347;
  M.~Spira, A.~Djouadi, D.~Graudenz and P.~M.~Zerwas in Ref.~\cite{gluon_fusion};
  M.~Kramer, E.~Laenen and M.~Spira,
  Nucl.\ Phys.\  B {\bf 511} (1998) 523
  [arXiv:hep-ph/9611272];
  A.~Djouadi, M.~Spira and P.~M.~Zerwas,
  Phys.\ Lett.\  B {\bf 264} (1991) 440;
  S.~Dawson,
  Nucl.\ Phys.\  B {\bf 359} (1991) 283;
  D.~Graudenz, M.~Spira and P.~M.~Zerwas,
  Phys.\ Rev.\ Lett.\  {\bf 70} (1993) 1372;
  R.~P.~Kauffman and W.~Schaffer,
  Phys.\ Rev.\  D {\bf 49} (1994) 551
  [arXiv:hep-ph/9305279];
  S.~Dawson and R.~Kauffman,
  Phys.\ Rev.\  D {\bf 49} (1994) 2298
  [arXiv:hep-ph/9310281].

\bibitem{QCD2}
  S.~Dawson, A.~Djouadi and M.~Spira,
  Phys.\ Rev.\ Lett.\  {\bf 77} (1996) 16
  [arXiv:hep-ph/9603423];
  M.~Muhlleitner and M.~Spira,
  Nucl.\ Phys.\  B {\bf 790} (2008) 1
  [arXiv:hep-ph/0612254].


\bibitem{Guasch:2001wv}
  J.~Guasch, W.~Hollik and S.~Penaranda,
  Phys.\ Lett.\ B {\bf 515} (2001) 367
  [arXiv:hep-ph/0106027].

\bibitem{three_body}
  W.~Y.~Keung and W.~J.~Marciano,
  Phys.\ Rev.\  D {\bf 30} (1984) 248;
  A.~Djouadi, J.~Kalinowski and P.~M.~Zerwas,
  Z.\ Phys.\  C {\bf 70} (1996) 435
  [arXiv:hep-ph/9511342];
  S.~Moretti and W.~J.~Stirling,
  Phys.\ Lett.\  B {\bf 347} (1995) 291
  [Erratum-ibid.\  B {\bf 366} (1996) 451]
  [arXiv:hep-ph/9412209];
  R.~Decker, M.~Nowakowski and A.~Pilaftsis,
  Z.\ Phys.\  C {\bf 57} (1993) 339
  [arXiv:hep-ph/9301283].


\bibitem{KL} I.B. Khriplovich and S.K. Lamoreaux, {\em CP Violation
  Without Strangeness} (Springer, New York, 1997).

\bibitem{PR} For a recent review, see, M.~Pospelov and A.~Ritz,
  Annals Phys.\  {\bf 318} (2005) 119.

\bibitem{CKP} For two-loop Higgs-mediated contributions to EDMs in the
  CP-violating MSSM, see D.  Chang, W.-Y. Keung and A. Pilaftsis,
Phys.\
  Rev.\ Lett.\ {\bf 82} (1999) 900; A. Pilaftsis, Nucl.\ Phys.\ B {\bf
  644} (2002) 263; D.~A.~Demir, O.~Lebedev, K.~A.~Olive, M.~Pospelov
and
  A.~Ritz, Nucl.\ Phys.\ B {\bf 680} (2004) 339;
  K.~A.~Olive, M.~Pospelov, A.~Ritz and Y.~Santoso,
  Phys.\ Rev.\  D {\bf 72} (2005) 075001
  [arXiv:hep-ph/0506106].

\bibitem{Ellis:2005ik}
  J.~R.~Ellis, J.~S.~Lee and A.~Pilaftsis,
  Phys.\ Rev.\ D {\bf 72} (2005) 095006
  [arXiv:hep-ph/0507046].

\bibitem{Lee:2007ai}
  J.~S.~Lee and S.~Scopel,
  Phys.\ Rev.\  D {\bf 75} (2007) 075001
  [arXiv:hep-ph/0701221];
%
  J.~S.~Lee,
  arXiv:0706.2222 [hep-ph].


\bibitem{Regan:2002ta}
  B.~C.~Regan, E.~D.~Commins, C.~J.~Schmidt and D.~DeMille,
  Phys.\ Rev.\ Lett.\  {\bf 88} (2002) 071805.

\bibitem{LEP_HIGGS}
  G.~Abbiendi {\it et al.}  [OPAL Collaboration],
  Eur.\ Phys.\ J.\ C {\bf 37} (2004) 49
  [arXiv:hep-ex/0406057];
  OPAL Physics Note PN505, {\tt
http://opal.web.cern.ch/Opal/pubs/physnote/html/pn505.html};
  A.~Heister {\it et al.}  [ALEPH Collaboration],
  Phys.\ Lett.\ B {\bf 526} (2002) 191
  [arXiv:hep-ex/0201014];
  J.~Abdallah {\it et al.}  [DELPHI Collaboration],
  Eur.\ Phys.\ J.\ C {\bf 32} (2004) 145
  [arXiv:hep-ex/0303013];
  P.~Achard {\it et al.}  [L3 Collaboration],
  Phys.\ Lett.\ B {\bf 545} (2002) 30
  [arXiv:hep-ex/0208042]; 
  S.~Schael {\it et al.}  [ALEPH Collaboration],
  Eur.\ Phys.\ J.\  C {\bf 47} (2006) 547
  [arXiv:hep-ex/0602042].


\bibitem{upsilon_visible}
  P.~Franzini {\it et al.},
  Phys.\ Rev.\ D {\bf 35}, 2883 (1987).


\bibitem{PDG}
  W.~M.~Yao {\it et al.}  [Particle Data Group],
  J.\ Phys.\ G {\bf 33} (2006) 1.

\bibitem{DP}
  A.~Dedes and A.~Pilaftsis,
  Phys.\ Rev.\  D {\bf 67} (2003) 015012
  [arXiv:hep-ph/0209306].

\bibitem{CDF:2007kv}
  T.~Aaltonen {\it et al.}  [CDF Collaboration],
  Phys.\ Rev.\ Lett.\  {\bf 100} (2008) 101802
  [arXiv:0712.1708 [hep-ex]].

\bibitem{HFAG}
  E.~Barberio {\it et al.}  [Heavy Flavor Averaging Group (HFAG)
                  Collaboration],
  arXiv:0704.3575 [hep-ex].

\bibitem{Carena:2000uj}
   M.~S.~Carena, D.~Garcia, U.~Nierste and C.~E.~M.~Wagner,
   Phys.\ Lett.\  B {\bf 499}, 141 (2001)
   [arXiv:hep-ph/0010003].
   G.~Degrassi, P.~Gambino and G.~F.~Giudice,
    JHEP {\bf 0012}, 009 (2000)
    [arXiv:hep-ph/0009337].

\bibitem{Borzumati:2004rd}
F.~Borzumati, J.~S.~Lee and W.~Y.~Song,
Phys.\ Lett.\ B {\bf 595} (2004) 347.
[arXiv:hep-ph/0401024].

\bibitem{Btaunu}
  K.~Ikado {\it et al.},
  Phys.\ Rev.\ Lett.\  {\bf 97} (2006) 251802.
  [arXiv:hep-ex/0604018];
  B.~Aubert  [The BABAR Collaboration],
  arXiv:0708.2260 [hep-ex].

\bibitem{BDTT}
  B.~Aubert {\it et al.}  [BABAR Collaboration],
  Phys.\ Rev.\ Lett.\  {\bf 96} (2006) 241802
  [arXiv:hep-ex/0511015].

\bibitem{Evans:2007hq}
  H.~G.~Evans  [CDF Collaboration],
  arXiv:0705.4598 [hep-ex].

\bibitem{b2sg-nnlo}
  M.~Misiak {\it et al.},
  Phys.\ Rev.\ Lett.\  {\bf 98} (2007) 022002,
  [arXiv:hep-ph/0609232].





\end{thebibliography}
\end{document}